\documentclass[fleqn,usenatbib]{mnras}

\usepackage[T1]{fontenc}

\DeclareRobustCommand{\VAN}[3]{#2}
\let\VANthebibliography\thebibliography
\def\thebibliography{\DeclareRobustCommand{\VAN}[3]{##3}\VANthebibliography}

\usepackage{graphicx}	
\usepackage{amsmath}	
\usepackage{amssymb}	
\usepackage{xcolor}

\title[Modeling Balmer line signatures of stellar CMEs]{Modeling Balmer line signatures of stellar CMEs}

\author[M. Leitzinger et al.]{
M. Leitzinger,$^{1}$\thanks{E-mail: martin.leitzinger@uni-graz.at}
P. Odert,$^{1}$
P. Heinzel$^{2,3}$
\\
$^{1}$Institute of Physics/IGAM, University of Graz, Universit\"atsplatz 5, 8010 Graz, Austria\\
$^{2}$Astronomical Institute, The Czech Academy of Sciences, 25165 Ond\v{r}ejov, Czech Republic\\
$^{3}$University of Wroc{\l}aw, Center of Scientific Excellence -- Solar and Stellar Activity, Kopernika 11, 51-622 Wroc{\l}aw, Poland}

\date{Accepted XXX. Received YYY; in original form ZZZ}

\pubyear{2015}

\begin{document}
\label{firstpage}
\pagerange{\pageref{firstpage}--\pageref{lastpage}}
\maketitle

\begin{abstract}
From the Sun we know that coronal mass ejections (CMEs) are a transient phenomenon, often correlated with flares. They have an impact on solar mass- and angular momentum loss, and therefore solar evolution, and make a significant part of space weather. The same is true for stars, but stellar CMEs are still not well constrained, although new methodologies have been established, and new detections presented in the recent past. So far, probable detections of stellar CMEs have been presented, but their physical parameters which are not directly accessible from observations, such as electron density, optical thickness, temperature, etc., have been so far not determined for the majority of known events. We apply cloud modeling, as commonly used on the Sun, to a known event from the literature, detected on the young dMe star V374~Peg. This event manifests itself in extra emission on the blue side of the Balmer lines. By determining the line source function from 1D NLTE modeling together with the cloud model formulation we present distributions of physical parameters of this event. We find that except for temperature and area all parameters are at the upper range of typical solar prominence parameters. The temperature and the area of the event were found to be higher than for typical solar prominences observed in Balmer lines. We find more solutions for the filament than for the prominence geometry. Moreover we show that filaments can appear in emission on dMe stars contrary to the solar case.
\end{abstract}




\section{Introduction}

The Sun and other stars are known to exhibit various magnetic activity phenomena, the most spectacular ones being outbreaks of radiation, so-called flares, and mass expulsions into the heliosphere, so-called coronal mass ejections (CMEs). Both are known for a long time from the Sun \citep{Gopalswamy2016,Benz2017}, and parameters of both phenmomena are statistically determined because of extensive solar monitoring during the last decades. Both phenomena are also known from stars, but not studied in such detail as on the Sun due to their larger distances. Flares on stars have been detected already in the first half of the last century \citep{Joy1949,Luyten1949} on L 726-8 or better known as UV~Ceti. Since then, especially with the exoplanet satellite missions Convection, Rotation and planetary Transits (CoRoT), Kepler, and now with the Transiting Exoplanet Survey Satellite (TESS), the number of detected flares was boosted to a new level. This allowed statistical investigations and because of that an improved understanding of flares on stars \citep[e.g.][]{Balona2015, Davenport2016, Guenther2020}. Moreover, a large number of so-called superflares (E>10$^{33}$erg) was also detected, enabling also a statistical analysis of this phenomenon for the first time \citep[e.g.][]{Maehara2012, Tu2020, Doyle2020}.\\
Stellar CME research is a younger branch of stellar astrophysics. As it is not possible to image stellar CMEs as it is done on the Sun, various signatures in different wavelength ranges have been used to attempt to detect this phenomenon on stars. The so far used  methodologies are a) Doppler-shifted emission/absorption as the direct signature of plasma moving away from the star, b) the stellar analogue of solar radio type II bursts being the signature of a shock wave (on the Sun often driven by CMEs), c) coronal dimmings being correlated to CMEs on the Sun, and d) X-ray absorptions during flares being caused by plasma obscuring the flaring region. Method a), Doppler-shifted emission/absorption, was used for main-sequence stars \citep{Houdebine1990, Gunn1994, Bond2001, FuhrmeisterSchmitt2004, Leitzinger2011a, Leitzinger2014, Vida2016, Korhonen2017, Vida2019, Leitzinger2020, Muheki2020a, Muheki2020b, Namekata2021}, for pre-main sequence stars \citep{Guenther1997}, and for giant stars \citep{Argiroffi2019}. Few distinctive events (events with larger Doppler-shift),  but many more possible events (candidate events with smaller Doppler-shifts) were detected from X-rays to the optical. Method b), radio type II bursts,  was used mainly on main-sequence M stars in different frequency domains \citep[from MHz to GHz;][]{Leitzinger2010, Boiko2012,Crosley2016, Crosley2018a, Crosley2018b, Villadsen2019}, but yet no stellar radio type II bursts were detected. Method c), coronal dimmings, has been recently successfully used on late-type main-sequence stars \citep{Veronig2021}, and more than 20 events were detected. Method d) has been used on RS CVn systems, weak-line T-Tauri stars as well as on M-type dwarfs \citep[][and references therein]{Moschou2019} and several such events have been detected. \\
Every method described above has its advantages, but also its drawbacks. Method a) is the only direct signature of ejected plasma, but it requires spectroscopic observations. It can be used very well in the optical, i.e. a wavelength domain which is easily accessible from the ground, in contrast to short wavelength domains such as UV or X-rays, which require satellite observations. On the other, hand this method suffers from the fact that the measured velocity is only the line-of-sight component. Method b) depends strongly on the sensitivity of the radio telescope, especially when assuming that type II bursts occur on stars also in the decameter/meter range where the sensitivity of existing radio facilities is very limited. Moreover, \citet{Mullan2019} suggested that CMEs on active stars may not necessarily drive shock waves and therefore may not cause type II bursts. Method c) is a very promising method and requires imaging only, and a well established relation of coronal dimmings and CMEs exists on the Sun \citep{Dissauer2019, Veronig2021}. On the other hand this method requires X-ray/EUV satellite observations which are harder to obtain than ground based optical observations. Finally, method d) requires also short wavelength observations and its interpretation requires additional modeling. \\
In the present study we focus on a better interpretation of optical spectroscopic observations (method a)) of possible stellar CMEs. This signature is typically thought to be caused by erupting filaments/prominences forming the dense core of CMEs. For this purpose we select the known event detected on the dMe star V374~Peg \citep{Vida2016} and apply the non-local thermal equilibrium (NLTE) approach (i.e. departures from local thermal equilibrium), as commonly applied on the Sun, to constrain the physical parameters of this event.\\


\section{Observations and modeling}
\subsection{The V374~Peg event}   
\begin{figure}
\begin{center}
	\includegraphics[width=8.2cm]{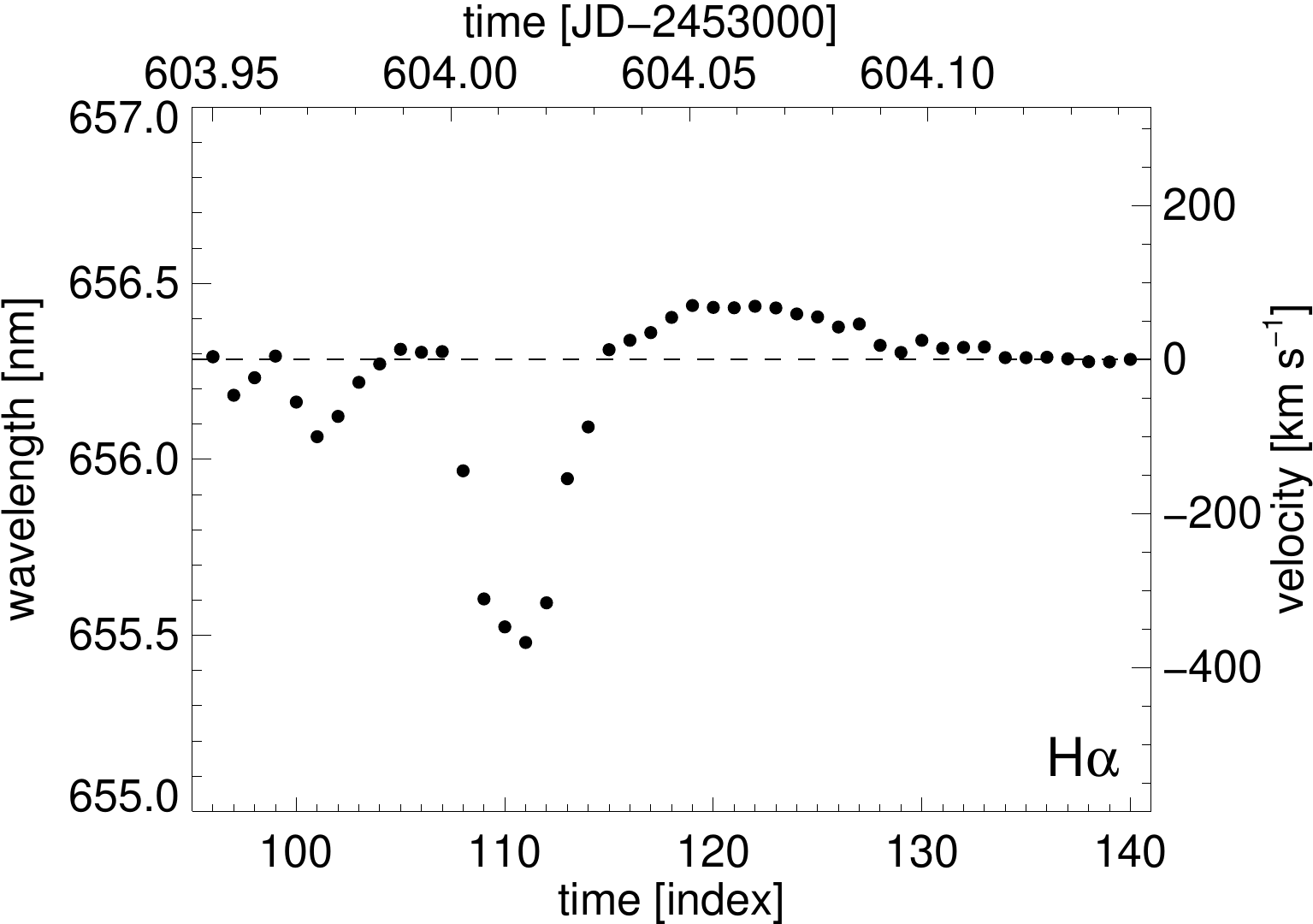}
	\includegraphics[width=8.2cm]{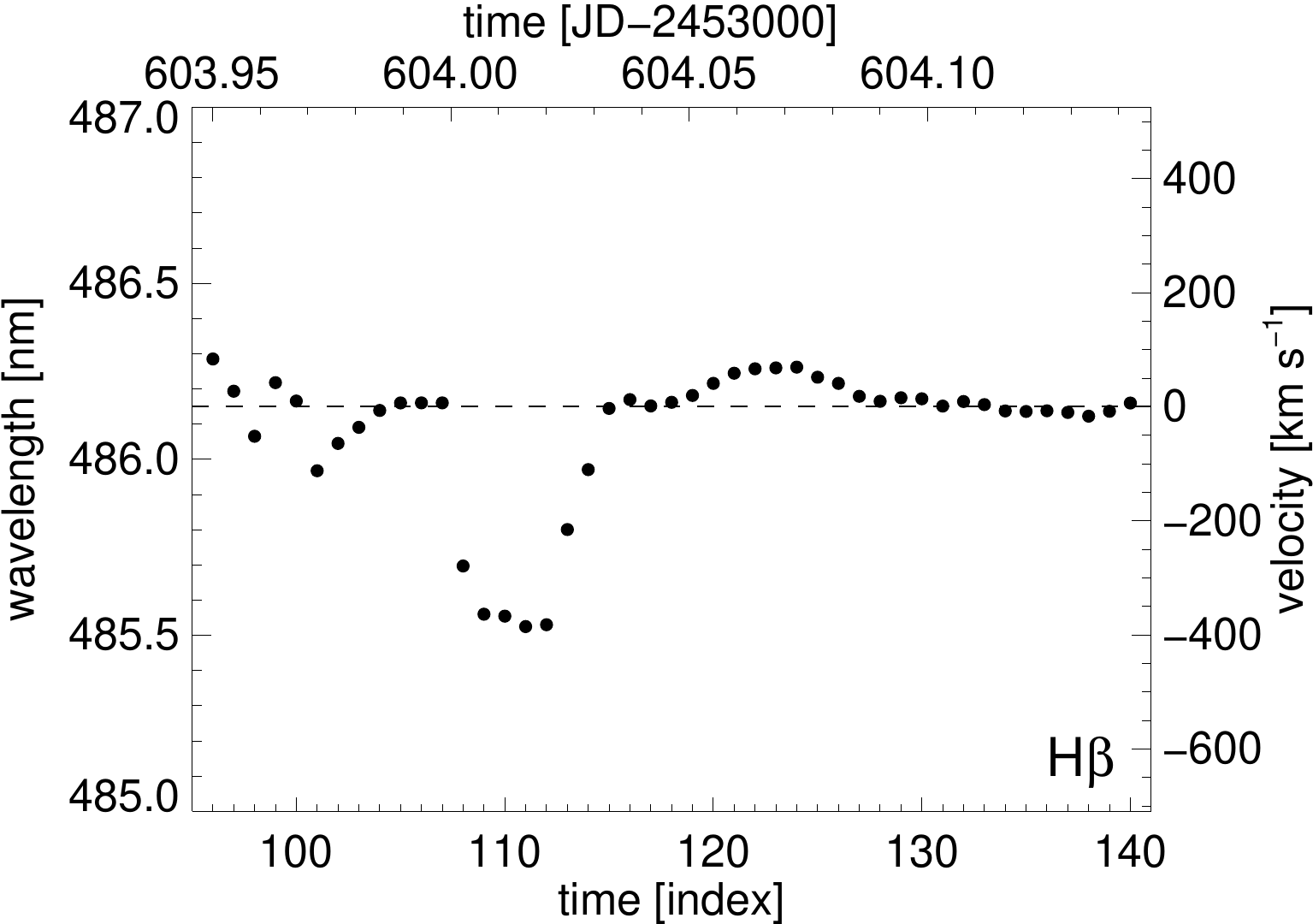}
    \caption{Line of sight bulk velocity in H$\alpha$ and H$\beta$ of the data set from the 21st of August 2005 covering a strong line asymmetry (spectra 109-112) on V374~Peg.\label{lineofsightplot}}
\end{center} 
\end{figure}

\begin{figure}
\begin{center}
	\includegraphics[width=9cm]{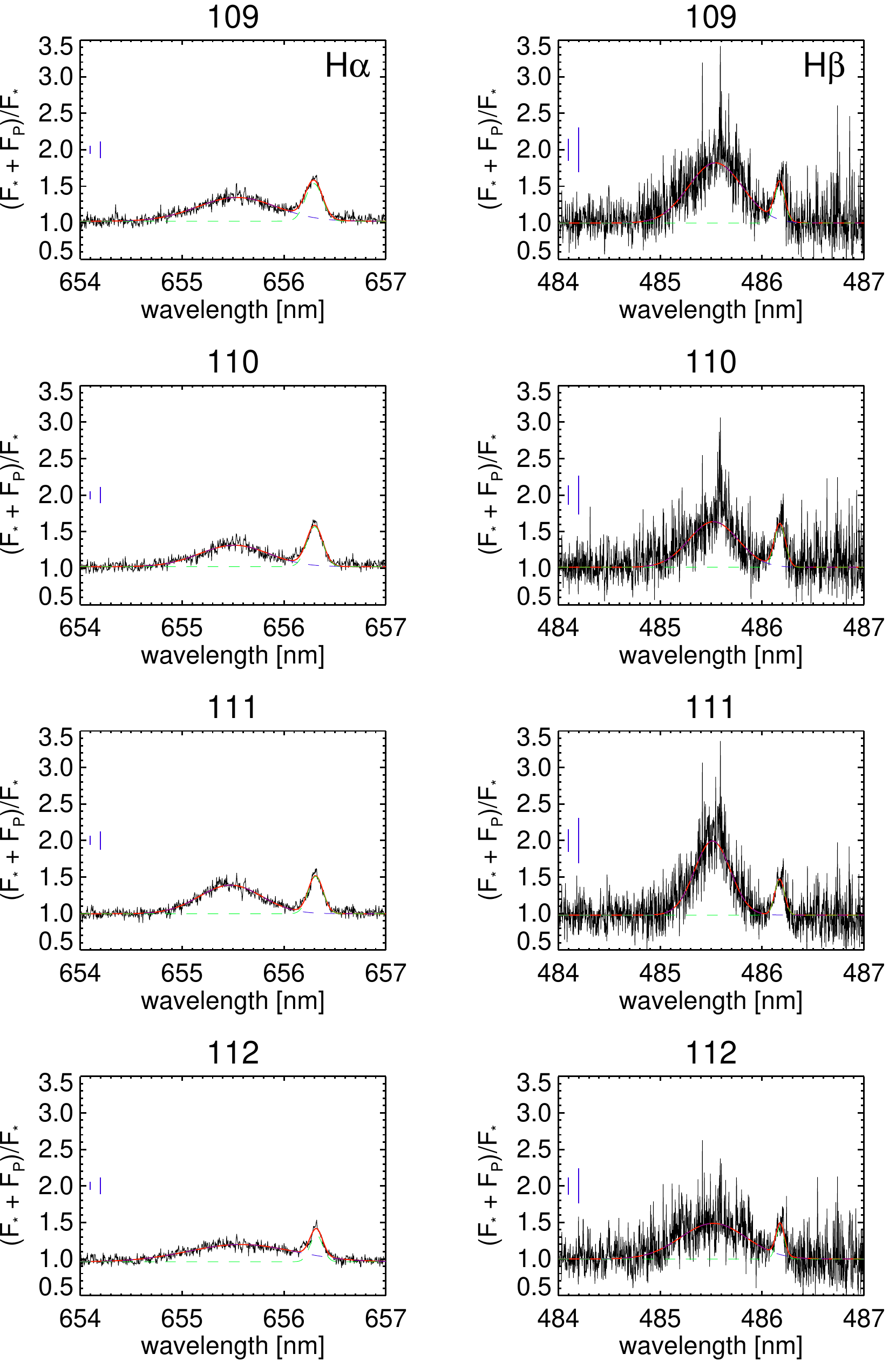}
    \caption{The four spectra of the event of interest, normalized by a quiescent spectrum, of V374~Peg showing the event to be modelled in this paper. Left column: Spectroscopic time series of the event in H$\alpha$ showing a distinct blue asymmetry. Right column: Spectroscopic time series of the event in H$\beta$ showing a distinct blue asymmetry. The event spectra are fitted by two Gaussians to take into account the evolution of the line core and the blue wing. Blue vertical lines indicate $\pm$ 1- and 2-$\sigma$ errors.\label{balmerplots}}
\end{center} 
\end{figure}

\citet{Vida2016} investigated the long-term activity of the fully convective dMe star V374~Peg. Beside other observations, data from the Canada-France-Hawaii-Telescope (CFHT) and its Echelle SpectroPolarimetric Device for the Observation of Stars (ESPaDOns) were analysed. The spectral resolution of the spectra shown in Fig.~\ref{balmerplots} around H$\alpha$ is $\sim$0.1\AA{} or $\sim$5~km~s$^{-1}$. The spectra were taken in spectropolarimetric mode and each spectrum covers the range of 3700-10400\AA{}. Observational characteristics of the spectra are given in table~\ref{peakfluxtable}. The spectra were obtained from the Polarbase archive\footnote{\url{http://polarbase.irap.omp.eu/}} and are available in normalized and non-normalized form. In these spectroscopic data several flares as well as blue- and red-wing asymmetries in Balmer lines have been detected. Among others, the most striking one showed a fast extra-emission occurring during flaring. This event was interpreted in \citet{Vida2016} as a possible CME. We focus in the present study on this event. The event was observed on the 21st of August 2005 revealing 2 flares with significant line asymmetries \citep[see Figs. 8, 12, and 13 in][]{Vida2016}. The first flare occurred with a strong symmetric broadening of the wings, the second weaker flare was only recognized by its line core enhancement and was followed by two subsequent blue extra emissions. Finally a long lasting extra-emission in the red followed.\\
In Fig.~\ref{lineofsightplot} we show the line of sight bulk velocity versus time for H$\alpha$ (left panel) and for H$\beta$ (right panel) for the blue and red extra emissions. To determine the velocities we fit the spectra with two Gaussian functions, one centered at the stellar Balmer lines, and the other representing the blue- and red-shifted line components (see Fig.~\ref{balmerplots}). The first blue extra-emission, possibly a small pre-event, is visible in spectra 100-102. The event of interest (second blue extra emission) starts already in spectrum 108 (rise of the blue-wing asymmetry) and disappears in spectrum 115. Then the line asymmetry switches to the red wing of H$\alpha$ already in spectrum 115 and lasts until spectrum 133 from where on the profile does not show an asymmetry anymore. For additional details see \citet{Vida2016}.\\
In Fig.~\ref{balmerplots} we show the H$\alpha$ and H$\beta$ event spectra normalized by a quiescent spectrum (single spectrum taken before the first strong flare), from which we determine the peak enhancement of the blue-shifted extra emissions (see table~\ref{peakfluxtable}). The blue dashed line marks the Gaussian fit to the extra emission, the green dashed line marks the Gaussian fit to the stellar line core, and the red solid line marks the double Gaussian fit to the total stellar profile. The two vertical blue lines indicate the $\pm$1- and 2-$\sigma$ error of the blue-shifted peak enhancement. The error represents the scatter of the data in the region of the peak enhancement around the fitted curve.

\begin{table*}
	\centering%
	\caption{Measured peak fluxes of the blue extra-emissions of the normalized spectra (see Fig.~\ref{balmerplots}) and observational characteristics.}%
	\label{peakfluxtable}
	\begin{tabular}{lcccccc} 
     index    &           JD          &        $(F_{\ast} + F_{P})/F_{\ast}$         &   $(F_{\ast} + F_{P})/F_{\ast}$        &        S/N        &       S/N      & t$_{\mathrm{exp}}$  \\
\hline
              &                       &               H$\alpha$                      &             H$\beta$                   &     H$\alpha$     &      H$\beta$  &         [s]            \\
\hline

no.109        &     2453604.00689     &            1.347 $\pm$ 0.058                 &             1.821 $\pm$ 0.153          &       46          &       19       &         300             \\  
no.110        &     2453604.01120     &            1.314 $\pm$ 0.056                 &             1.637 $\pm$ 0.132          &       46          &       19       &         300             \\
no.111        &     2453604.01552     &            1.389 $\pm$ 0.063                 &             1.995 $\pm$ 0.157          &       45          &       19       &         300             \\
no.112        &     2453604.01984     &            1.198 $\pm$ 0.057                 &             1.487 $\pm$ 0.119          &       45          &       18       &         300             \\
	\hline
	\end{tabular}
\end{table*}

\subsection{Modeling}
\begin{figure}
\begin{center}
    \includegraphics[width=8cm]{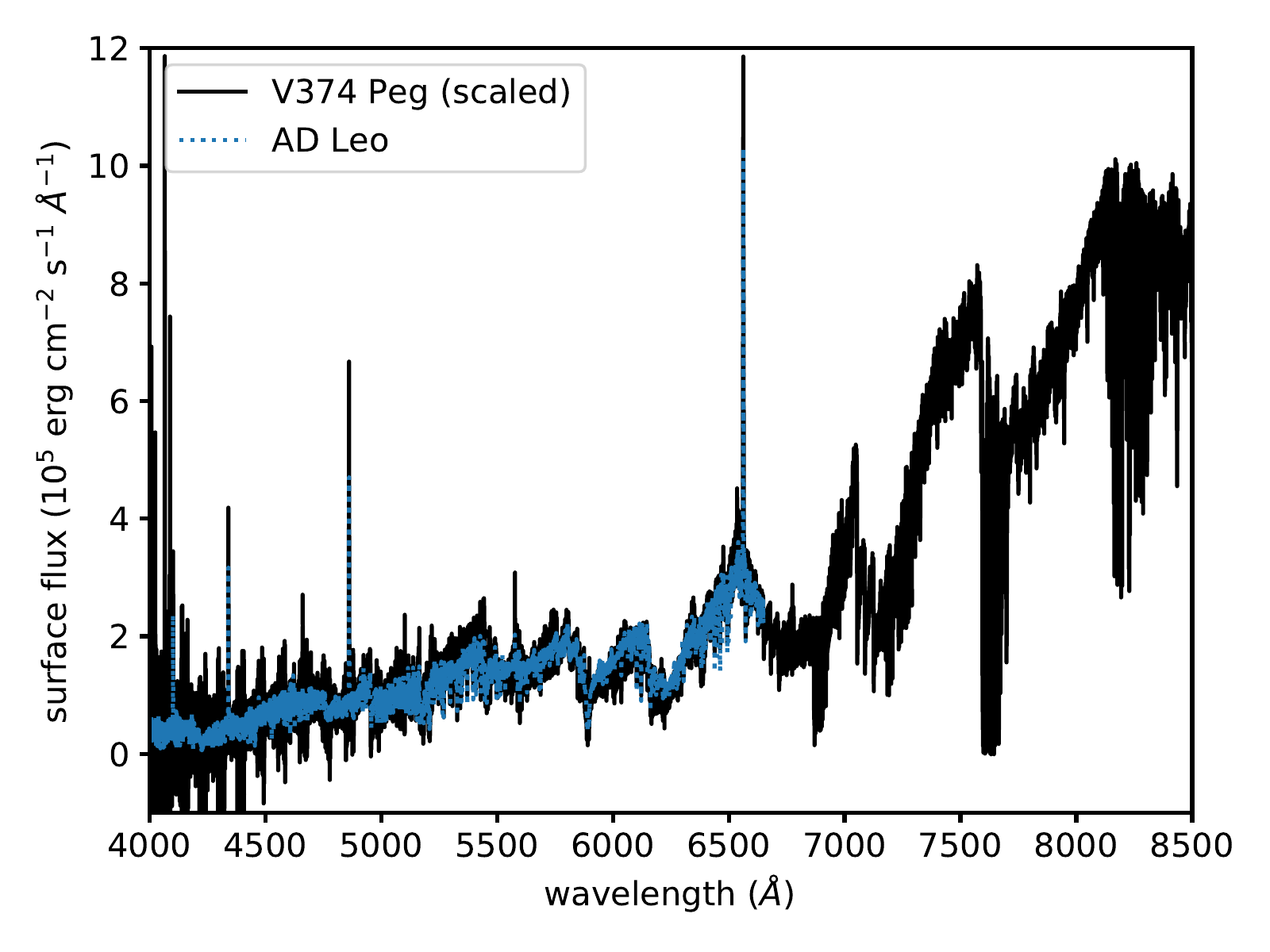}
    \caption{Optical spectra of AD~Leo (light blue dotted line) and V374~Peg (black solid line). As one can see, the optical continua are similar and Balmer line peaks of V374~Peg are somewhat higher than for AD~Leo.\label{V374PegADLeo}}
\end{center} 
\end{figure}

We apply a cloud model formalism \citep{Beckers1964}, typically used for solar filaments, to these observations to deduce parameters of the expelled plasma. To reproduce the observed extra-emissions we consider two possible cases, namely filament and prominence geometries. In the filament case, the plasma is located in front of the stellar disk, whereas in the prominence case the plasma is located aside from the stellar disk. What we measure on a star during an eruptive event is always the sum of the stellar radiation and the flux from the   filament/prominence \citep[see][]{Odert2020}, shifted by its line of sight velocity. Accordingly we can write for the normalized (by the quiet pre-event spectrum) flux in the prominence case:\\
\begin{equation}
F = \frac{F_{\ast} + F_{P}}{F_{\ast}} = \frac{I_{\ast}A_{\ast} + I_{P}A_{P}}{I_{\ast}A_{\ast}} = \frac{S}{I_\ast} \frac{A_{P}}{A_{\ast}}(1 - e^{-\tau}) +1
\label{eqprom}
\end{equation}
by using 
\begin{equation}
I_{P} = S(1 - e^{-\tau}).
\end{equation}
For the filament case we need to write
\begin{equation}
\begin{split}
F &= \frac{F_{\ast} + F_{P}}{F_{\ast}} = \frac{I_{\ast}(A_{\ast} - A_{P}) + I_{P}A_{P}}{I_{\ast}A_{\ast}} \\
&= (\frac{S}{I_\ast} - 1) \frac{A_{P}}{A_{\ast}}(1 - e^{-\tau}) +1
\end{split}
\label{eqfil}
\end{equation}
by using
\begin{equation}
I_{P} = I_{\ast}e^{-\tau} + S(1 - e^{-\tau}).
\end{equation}
Here, $F_{\ast}$ is the stellar flux, $F_{P}$ is the prominence flux, $I_{\ast}$ is the stellar intensity, $I_{P}$ is the prominence intensity, and $A_{\ast}$ is the stellar disk area \citep[$A_{\ast}=\pi R_{\ast}^{2}$, where we adopt $R_{\ast}=0.3242R_{\sun}$;][]{Cifuentes2020}. Eq.~2 and Eq.~4 are simplified solutions to the radiative transfer equation \citep[see e.g.][]{Labrosse2010, Heinzel2015}, assuming a constant line source function and an emission normal to the slab representing the filament/prominence. For observed values of $(F_{\ast} + F_{P})/F_{\ast}$ of the V374~Peg event see table~\ref{peakfluxtable}.\\
The unknowns in Eq.~\ref{eqprom} and Eq.~\ref{eqfil} are the line source function $S$, the prominence area $A_{P}$, and its optical thickness $\tau$. To determine the line source function and the optical thickness of the prominence material we use an NLTE hydrogen model \citep{Heinzel1995, Heinzel1999} which was and is extensively used for solar filaments and prominences \citep[e.g.][]{Heinzel2014, Heinzel2016}, both quiescent and eruptive. The NLTE model approximates filaments/prominences as a 1-D slab \citep[see][]{Heinzel1999}.\\ 
The V374~Peg event was detected in H$\alpha$, H$\beta$, H$\gamma$, and H$\delta$ \citep[see][]{Vida2016}. For the present study we use only H$\alpha$ and H$\beta$ because the higher Balmer lines have much more noise and the NLTE model is a five level atom and modeling of the uppermost level for H$\gamma$ involves larger uncertainties.\\
The NLTE code needs as input the stellar spectral line profiles of the hydrogen Lyman, Balmer, and Paschen series. This is the radiation incident upon the prominence/filament slab, defining the boundary conditions for the radiative-transfer equation. To our knowledge there are no spectroscopic observations of the far ultra-violet (FUV) and near infra-red (NIR) of V374~Peg, and there is no flux calibrated spectrum of this star in the optical. Therefore we adopt a spectrum of AD~Leo (see Fig.~\ref{V374PegADLeo}), which is a star of similar age (V374Peg: 200~Myr \cite{Montes2001}, ADLeo: 25-300~Myr \citet{Shkolnik2009} and spectral type \citep[AD Leo: M3, V374Peg: M4;][]{Reid1995}, to flux-calibrate the CFHT spectra of V374~Peg. In Fig.~\ref{V374PegADLeo} we show the spectrum of AD~Leo (light-blue dotted line) which was taken from a library of flux-calibrated Echelle spectra \citep{Cincunegui2004} and the scaled CFHT spectrum of V374~Peg (black solid line). We convert the flux-calibrated spectrum of AD Leo to surface flux using its distance and radius \citep[$d=4.966$~pc; $R=0.4681R_{\sun}$;][]{Cifuentes2020}. Then we compute a scaling factor for the V374~Peg spectrum by minimizing the offset between the spectra, thereby assuming that the surface fluxes of both stars are similar based on their similar spectral types and effective temperatures \citep[AD Leo: 3300~K, V374Peg: 3200~K;][]{Cifuentes2020}. The excellent agreement between both spectra can be seen in Fig.~\ref{V374PegADLeo}. 
For the UV and FUV we use observations of AD Leo by the Hubble Space Telescope (HST) with its Space Telescope Imaging Spectrograph (STIS) and the Far Ultra-violet Spectroscopic Explorer (FUSE) to cover the Ly$\alpha$ (covered by HST/STIS), Ly$\beta$ (covered by FUSE), and Ly$\gamma$ (covered by FUSE) spectral lines. We assume again that the surface flux of AD~Leo is comparable to the surface flux of V374~Peg. As the Ly$\alpha$ line is affected by interstellar absorption, we adopt the reconstructed intrinsic line profile from \citet{Wood2005b}. The FUSE data are flux-calibrated and are taken from the MAST archive\footnote{\url{https://archive.stsci.edu/}}. \\
The spectral line and continuum intensities needed for the NLTE code (Paschen, Brackett) need to be approximated as no observations of those are available for V374~Peg and for AD~Leo.
The level of the Paschen and Brackett spectral lines and continua is determined by using the intensity computed for the effective temperature of V374~Peg and at the frequency of the line or continuum. For the Lyman continuum we assumed a detailed radiative balance (i.e. the radiative rates of the photoionisation and recombination are balanced) because also here no observations are available. The assumption of detailed radiative balance was justified by NLTE test runs for the solar case as here the incident intensities of the Lyman continuum are known.



\begin{figure*}
\begin{center}
	\includegraphics[width=8cm]{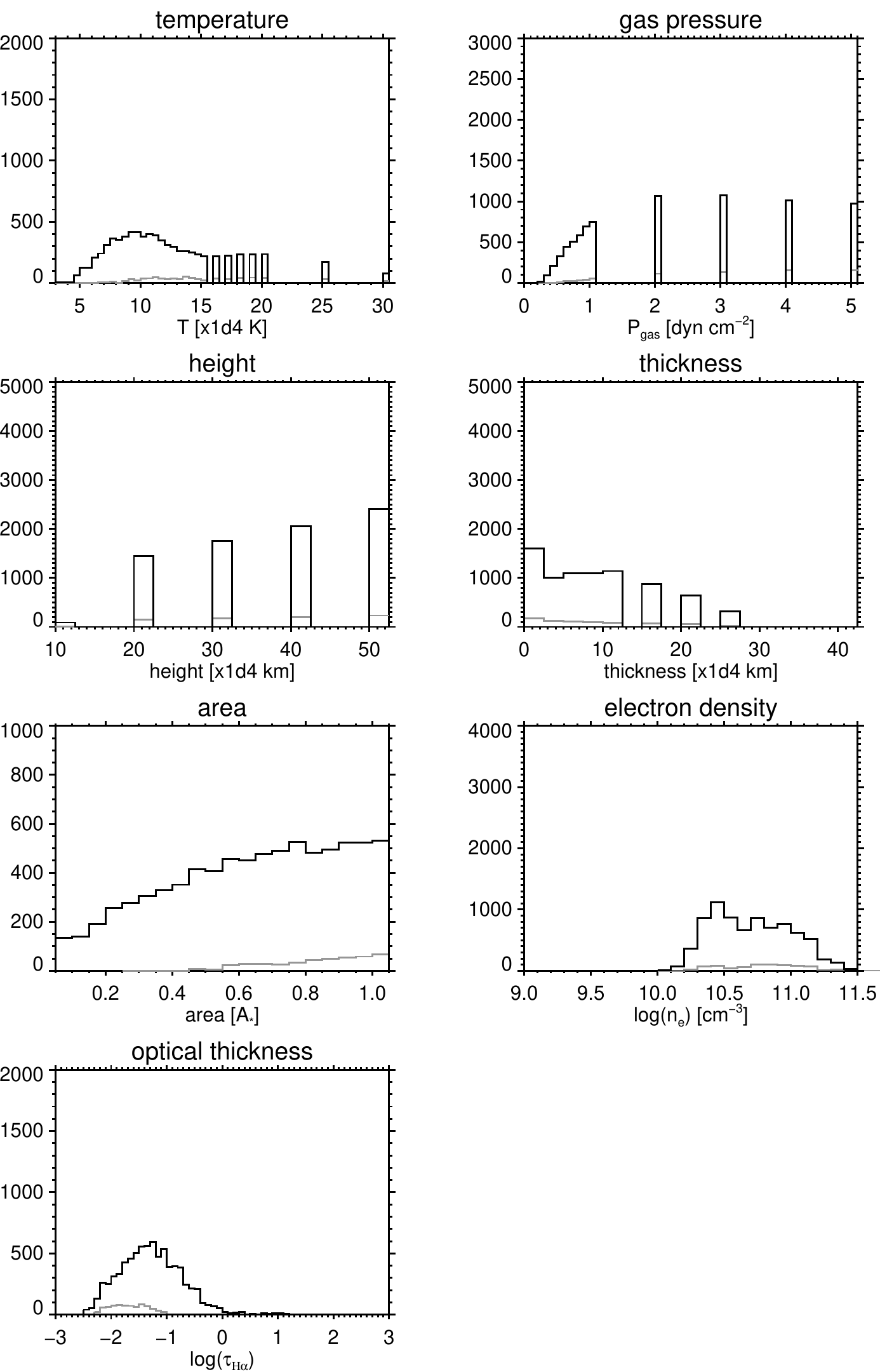}
	\includegraphics[width=8cm]{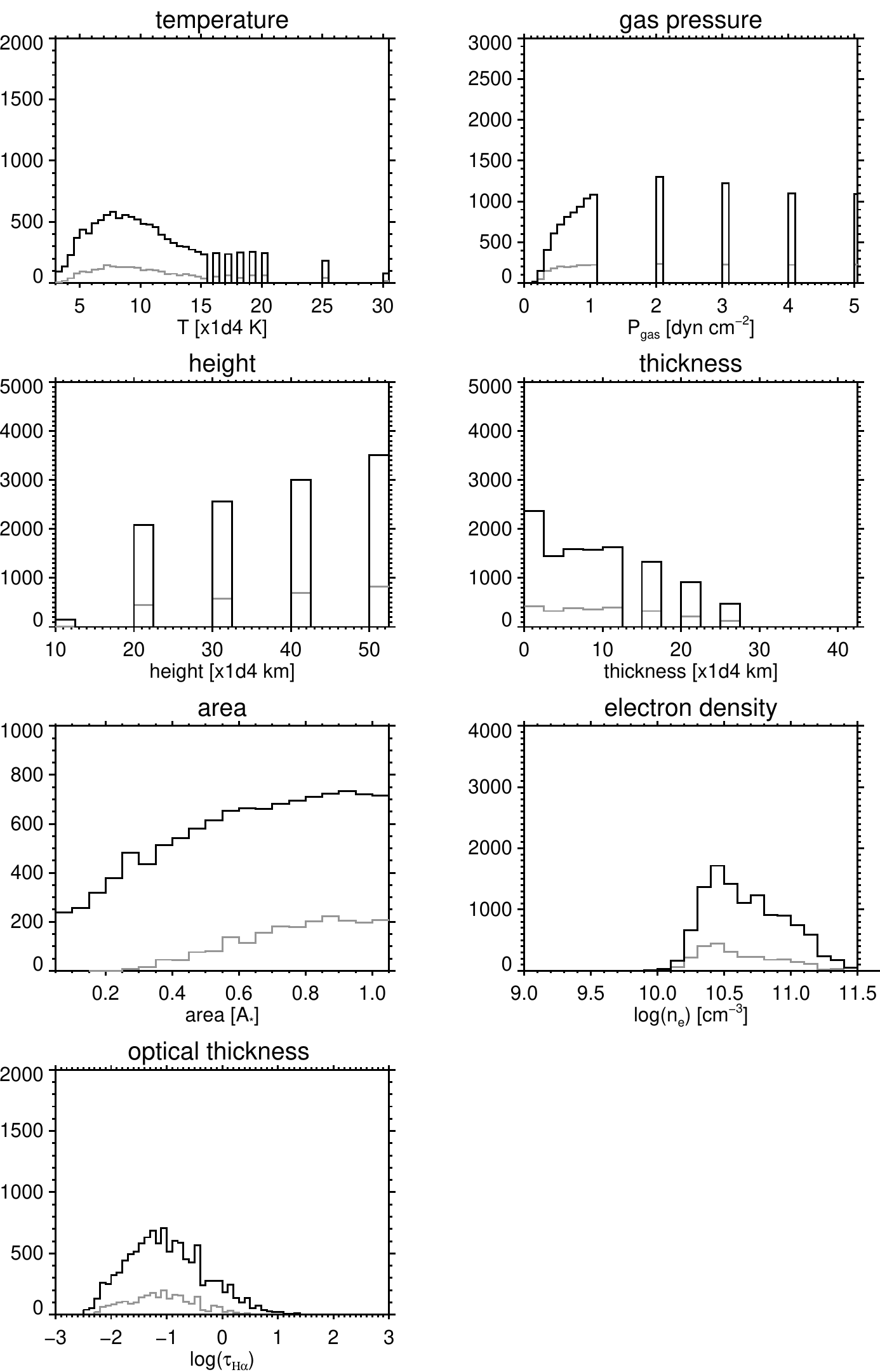}
    \caption{Histograms of the parameters of the cloud model results for spectrum 109. Columns 1 and 2: Histograms for the prominence case. Columns 3 and 4: Histograms for the filament case. Black solid lines refer to 2-$\sigma$ results whereas grey solid lines refer to 1-$\sigma$ results.\label{histograms1d238}}
\end{center} 
\end{figure*}
\section{Results}
\label{modelling}
Before starting modeling we need to define the input parameter grids. The NLTE code demands the following parameters: gas pressure, temperature, prominence height, prominence geometrical thickness, turbulent velocity, and radial velocity of the erupting prominence relative to the star. We used a gas pressure grid ranging from 0.01-5~dyn~cm$^{-2}$, a temperature grid ranging from 5000-300000~K, a height grid of 10000-500000~km (0.44-2.22~R$_{\ast}$), and a thickness grid ranging from 1000-400000~km (0.0044-1.77~R$_{\ast}$). Turbulent and radial velocities are fixed at 5 and 400~km~s$^{-1}$ (corresponding to the observed line of sight velocity), respectively. Additionally, for the area ratio ($A_{P}$/$A_{\ast}$) entering Eq.~\ref{eqprom} and \ref{eqfil}, we use a grid from 0.05-1.0. The input grids are not evenly spaced. In total we run 61568 parameter combinations.\\
The NLTE code provides the source function $S$ for the Balmer lines, their line center optical thicknesses $\tau$, and the electron density. The line source function is the same for both prominence and filament cases, distinction between the two structures is made only in the cloud model depending on the viewing direction. We match the observed peak fluxes of the blue extra-emissions with the modelled ones (see Eq.~\ref{eqprom}, ~\ref{eqfil}) within $\pm$ 1- and 2-$\sigma$ of the observed values, simultaneously for H$\alpha$ and H$\beta$. Doing so yields the parameter histograms shown in Fig.~\ref{histograms1d238} for spectrum no.~109. We produce such histograms for spectra 109-112 (for histograms of spectra 110-112 see Fig.~\ref{histograms1d239},~\ref{histograms1d240},~\ref{histograms1d241} in the appendix) as those spectra showed the fastest bulk velocities (cf. Fig.~\ref{lineofsightplot}). The first two columns in Fig.~\ref{histograms1d238} show the parameter histograms for the prominence geometry (see Eq.~\ref{eqprom}) and the last two columns show the parameter histograms for the filament geometry (see Eq.~\ref{eqfil}). For both prominence and filament geometry we show the parameter histograms for the 2-$\sigma$ (black solid line) and 1-$\sigma$ (grey solid line) results, i.e. cases where $F_{obs}$-$2\sigma$ $<$ $F_{NLTE}$ $<$ $F_{obs}$+$2\sigma$ and $F_{obs}$-$1\sigma$ $<$ $F_{NLTE}$ $<$ $F_{obs}$+$1\sigma$. Moreover we removed cases where thickness is greater than height, to avoid non-physical scenarii where the structure would merge with the star in the filament case. The parameter histograms for spectra 110-112 are shown in the appendix (see Fig.~\ref{histograms1d239},~\ref{histograms1d240},~\ref{histograms1d241}). We want to note that the number of possible cases is lowest for spectrum 111 and therefore the significance of the parameter histogram of this spectrum/event is of low significance. The bins of the 1D and 2D parameter histograms presented in the following are no central bins, i.e. refer to the lower bin edges. The parameter histograms of all four spectra, for both prominence and filament cases, indicate the following for the 2-$\sigma$ results (see also table~\ref{paramvalues}):\\
\textit{Temperature:} The median for the prominence case ranges from 105000-115000~K whereas the median for the filament case is somewhat lower and ranges from 95000-100000~K. The temperature distribution for all spectra and cases is broad, which is evident from the percentiles (10 and 90\%).  \\
\textit{Gas pressure:} The prominence case reveals a median in the range of 1-2~dyn~cm$^{-2}$, whereas the filament case shows a median in the range of 0.9-1~dyn~cm$^{-2}$. \\
\textit{Height:} Both, prominence and filament cases, show the same result, namely a median of 400000~km (1.77~R$_{\ast}$).\\
\textit{Thickness:} Both, prominence and filament cases, show a median of 75000~km (0.33~R$_{\ast}$) for the first three spectra. The last spectrum shows a median of 50000~km (0.22~R$_{\ast}$) for both prominence and filament cases.\\
\textit{Area:} The median for the prominence and filament cases ranges from 0.55-0.7~A$_{P}$/A$_{\ast}$ .\\
\textit{Electron density:} The distribution for logarithmic electron density shows for both prominence and filament cases median values in the range of 10.6-10.7.\\
\textit{Optical thickness:} Both, prominence and filament cases, show small values for $\tau_{H_{\alpha}}$ in the range of 0.03-0.05 for the prominence case, and 0.04-0.1 for the filament case.\\
\begin{figure*}
\begin{center}
	\includegraphics[width=7.3cm]{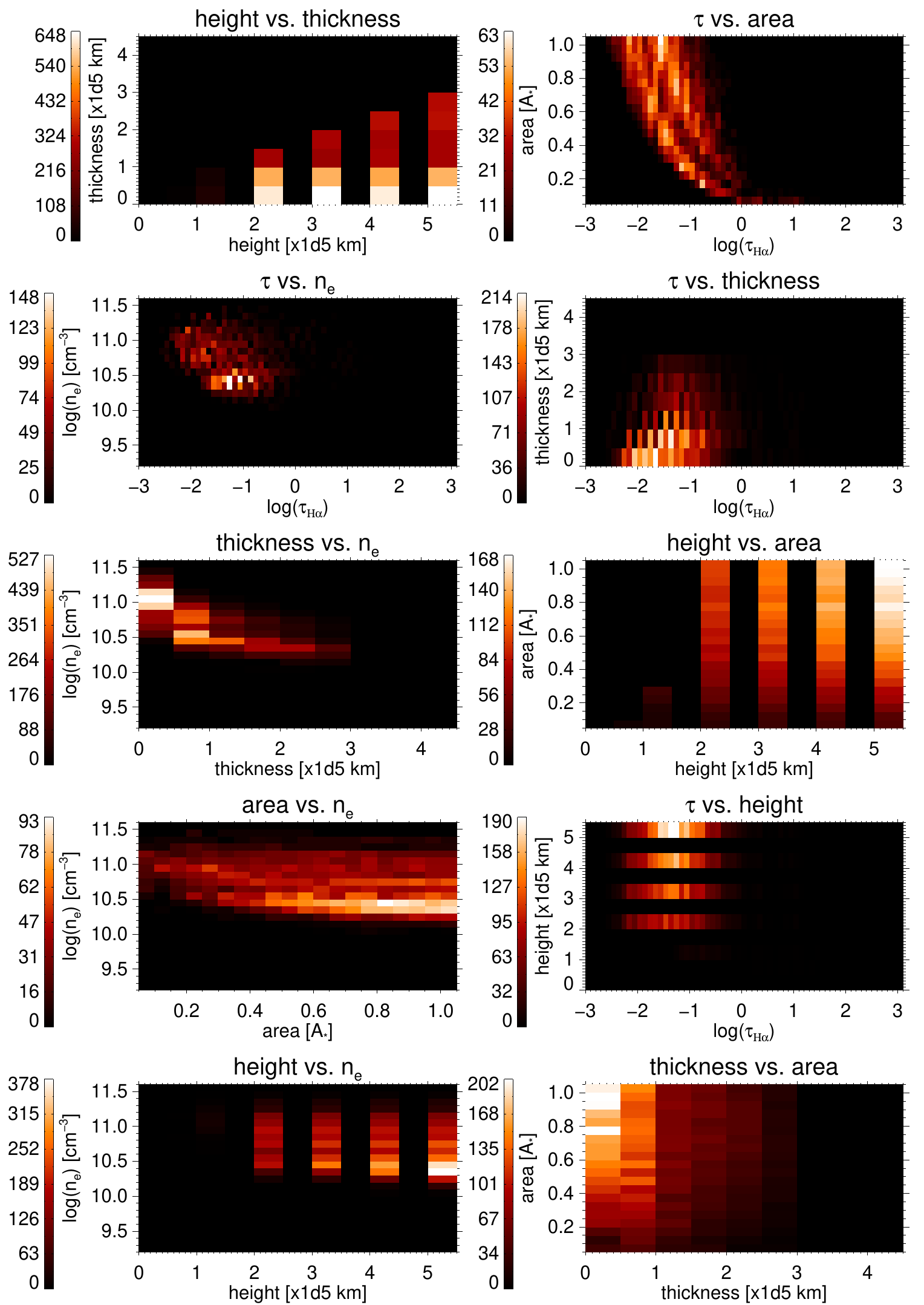}
	\includegraphics[width=7.3cm]{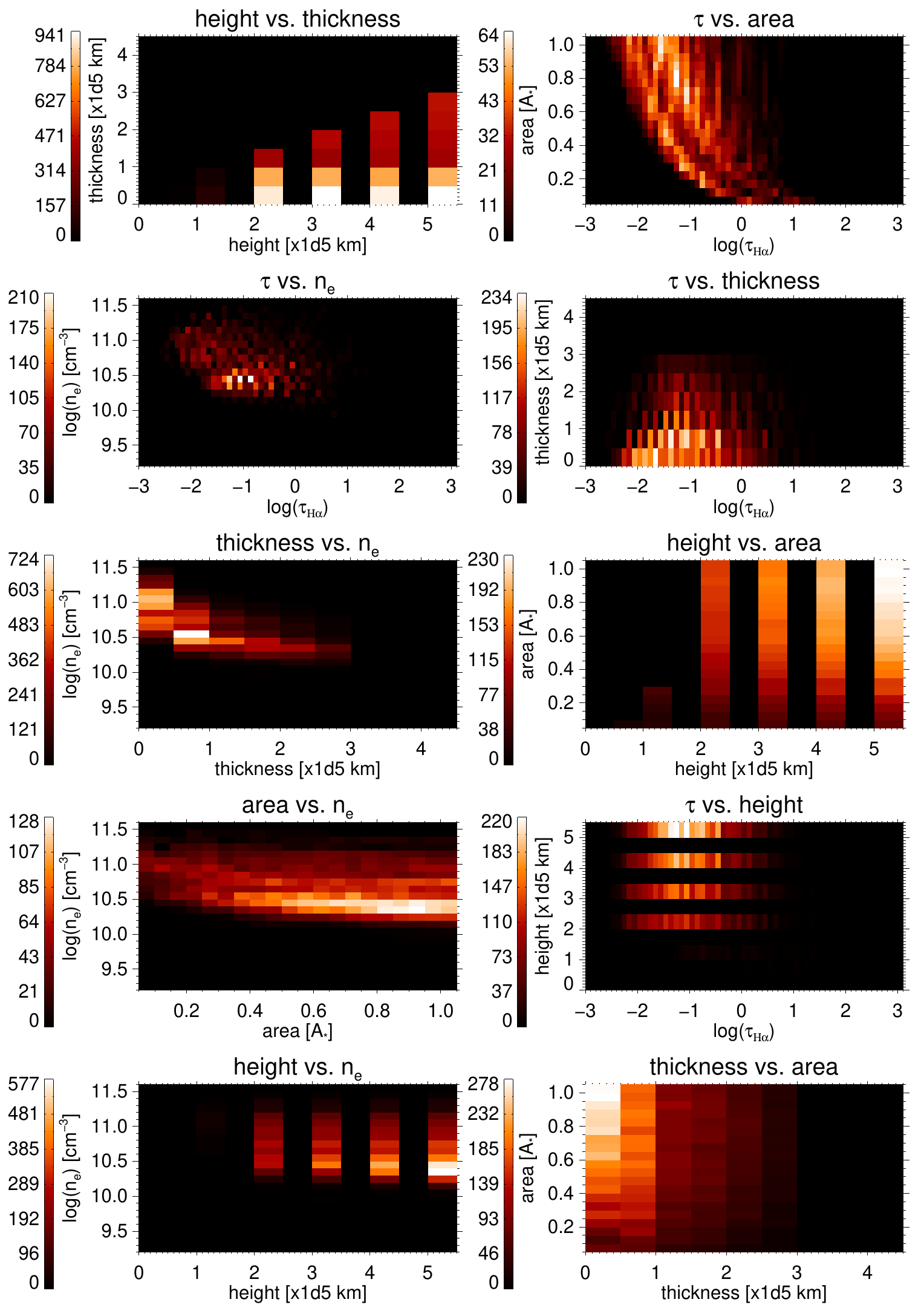}
	\includegraphics[width=7.3cm]{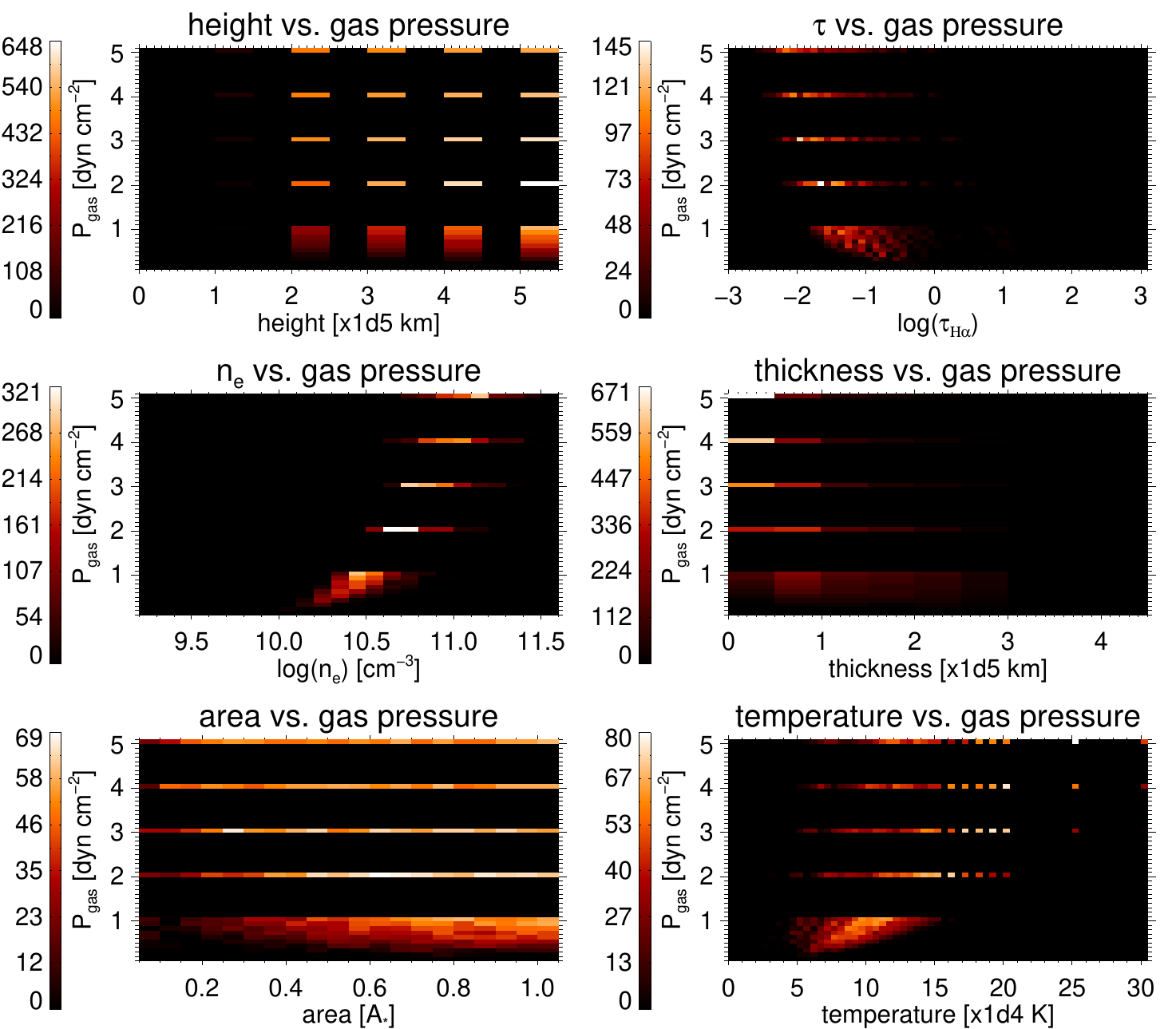}
	\includegraphics[width=7.3cm]{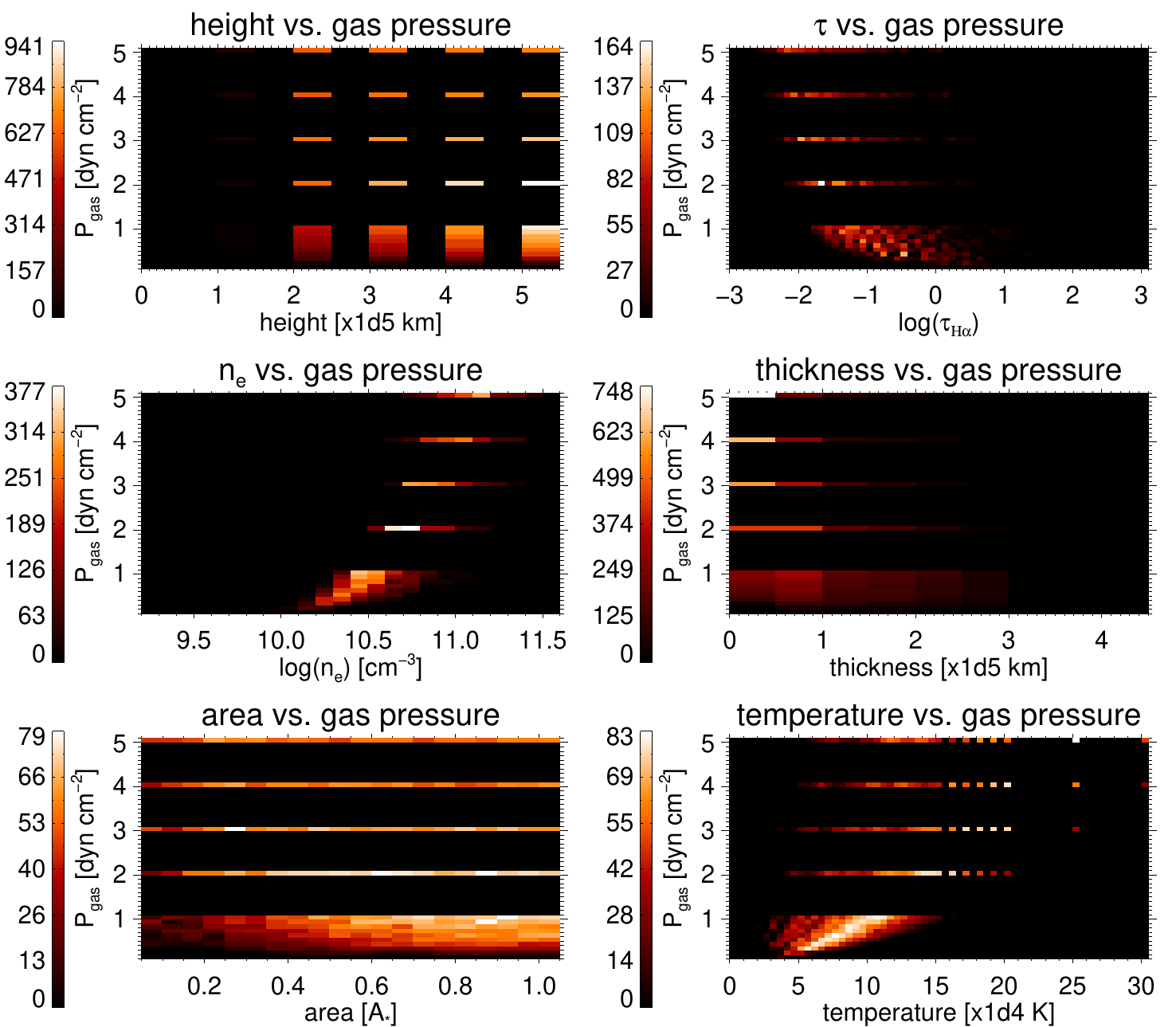}
	\includegraphics[width=7.3cm]{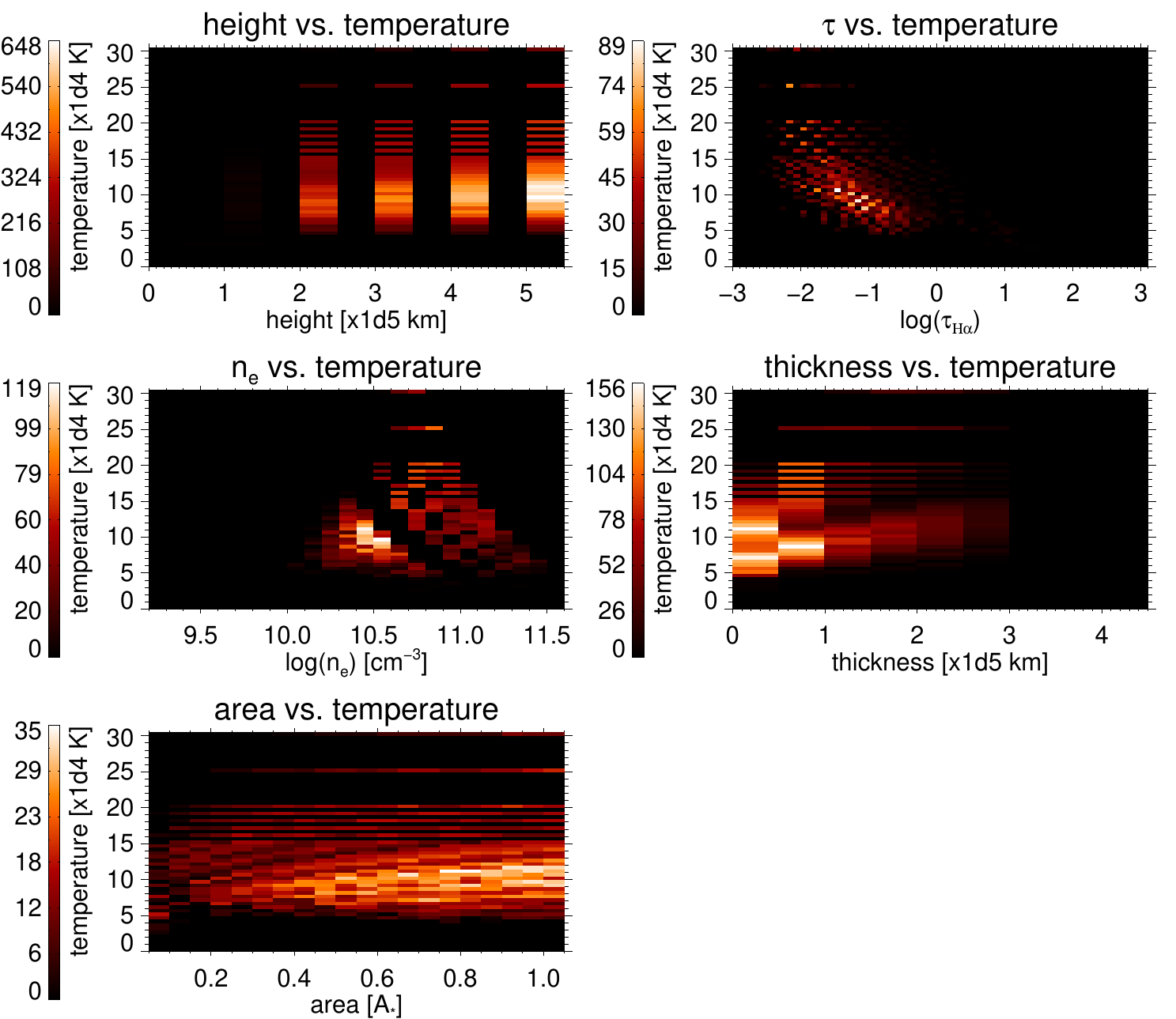}
    \includegraphics[width=7.3cm]{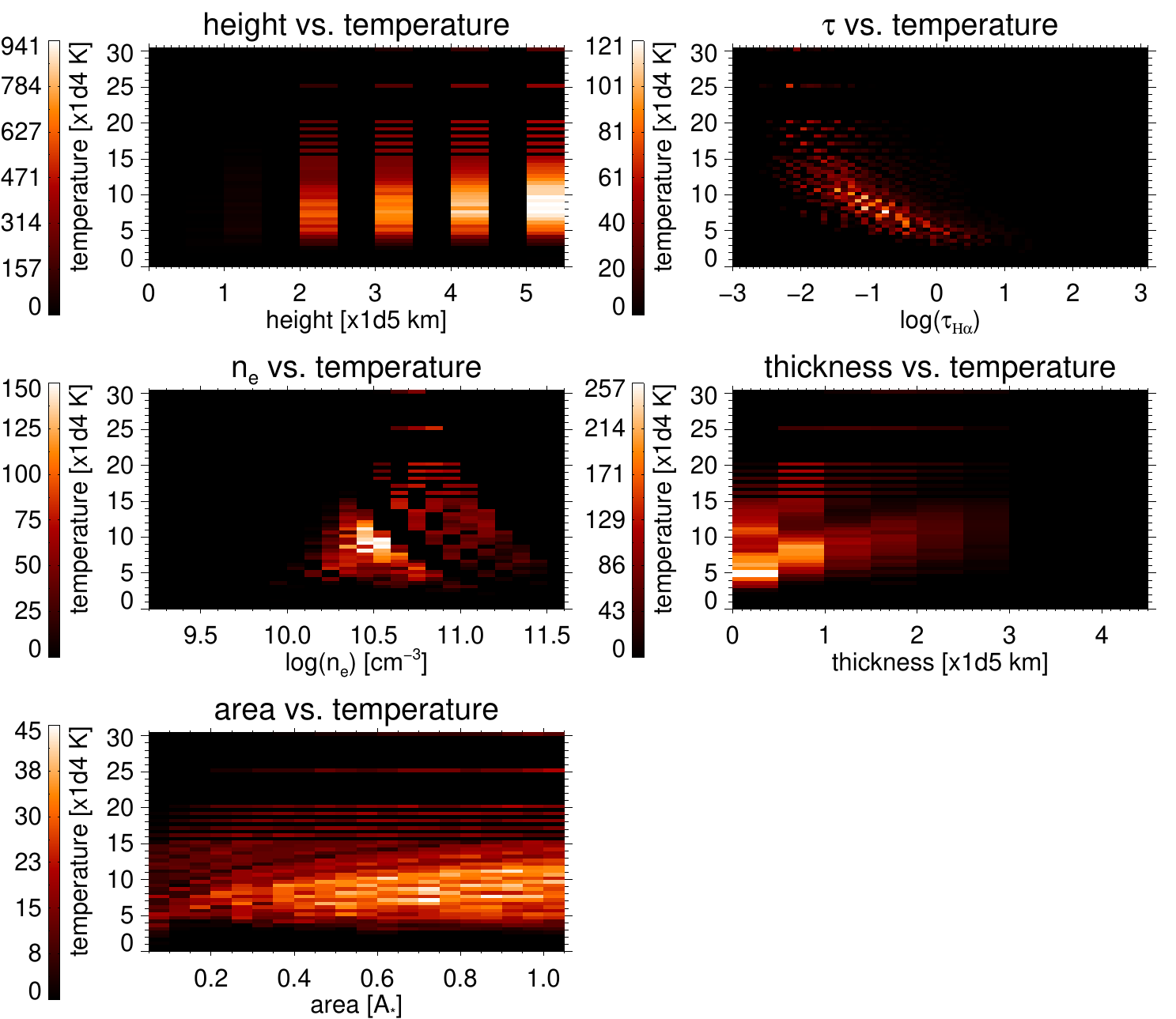}
    \caption{2D histograms of all parameter combinations of the 2-$\sigma$ cloud model results for spectrum no.~109. Columns 1 and 2: 2D histograms of all parameter combinations of the 2-$\sigma$ cloud model results for the prominence case. Columns 3 and 4: 2D histograms of all parameter combinations of the 2-$\sigma$ cloud model results for the filament case.\label{dep1}}
\end{center}
\end{figure*}
\begin{table*}
	\centering%
	\caption{Number of matches and resulting median values with percentiles (10 and 90\%) of the parameters (2-$\sigma$ results), as well as maximum bins of the histograms. Logarithmic electron density is given in cgs units.}%
	\label{paramvalues}
	\begin{tabular}{lcccc} 
\hline
              &                no.109               &                   no.110            &             no.111                  &        no.112                         \\
\hline
filament case &                                     &                                     &                                     &                                      \\
\hline
matches (1/2 $\sigma$) &            2545/11321               &               6653/16595            &             410/6734                &       6322/20426         \\
temperature [K]   & 95000$^{+75000}_{-45000}$/75000    & 95000$^{+75000}_{-45000}$/75000     & 95000$^{+75000}_{-40000}$/85000     & 100000$^{+70000}_{-50000}$/80000      \\
gas pressure [dyn~cm$^{-2}$]  & 1.0$^{+3.0}_{-0.6}$/2.0             &     1.0$^{+3.0}_{-0.5}$/2.0         &   1.0$^{+4.0}_{-0.6}$/2.0           & 0.9$^{+3.1}_{-0.5}$/2.0   \\
height [km]       & 400000$^{+100000}_{-200000}$/500000 & 400000$^{+100000}_{-200000}$/500000 & 400000$^{+100000}_{-200000}$/500000 & 400000$^{+100000}_{-200000}$/500000   \\
thickness [km]    & 75000$^{+125000}_{-70000}$/$<$25000 & 75000$^{+125000}_{-70000}$/$<$25000 & 75000$^{+125000}_{-65000}$/$<$25000 & 50000$^{+150000}_{-45000}$/$<$25000   \\
area [A$_{\ast}$]          & 0.6$^{+0.35}_{-0.40}$/0.9           &    0.6$^{+0.35}_{-0.40}$/0.8        &  0.7$^{+0.25}_{-0.40}$/0.95         & 0.55$^{+0.35}_{-0.40}$/0.6 \\
log(n$_{e}$)         & 10.6$^{+0.46}_{-0.30}$/10.4         &  10.6$^{+0.47}_{-0.29}$/10.4        &  10.6$^{+0.46}_{-0.30}$/10.4        & 10.6$^{+0.48}_{-0.31}$/10.3  \\
$\tau_{H\alpha}$        & 0.09$^{+0.85}_{-0.07}$/0.08         &    0.07$^{+0.84}_{-0.06}$/0.05      &  0.10$^{+0.65}_{-0.09}$/0.08        & 0.04$^{+0.62}_{-0.04}$/0.02  \\
\hline
prominence case&                                    &                                     &                                     &                                     \\
\hline
matches (1/2 $\sigma$)&             758/7767               &           4110/13773                &         12/3926                     &       3668/16564     \\
temperature [K]  & 110000$^{+70000}_{-45000}$/90000    & 105000$^{+75000}_{-45000}$/95000    & 115000$^{+75000}_{-40000}$/95000    & 110000$^{+70000}_{-50000}$/100000\\
gas pressure [dyn~cm$^{-2}$] & 2.0$^{+3.0}_{-1.4}$/3.0             &     1.0$^{+4.0}_{-0.5}$/2.0         &   2.0$^{+3.0}_{-1.3}$/3.0           & 1.0$^{+4.0}_{-0.5}$/2.0\\
height [km]       & 400000$^{+100000}_{-200000}$/500000 & 400000$^{+100000}_{-200000}$/500000 & 400000$^{+100000}_{-200000}$/500000 & 400000$^{+100000}_{-200000}$/500000\\
thickness [km]    & 75000$^{+125000}_{-70000}$/$<$25000 & 75000$^{+125000}_{-70000}$/$<$25000 & 75000$^{+125000}_{-65000}$/$<$25000 & 50000$^{+150000}_{-45000}$/$<$25000\\
area [A$_{\ast}$]          & 0.65$^{+0.30}_{-0.40}$/1.0          &    0.6$^{+0.35}_{-0.40}$/0.7        &  0.7$^{+0.25}_{-0.40}$/0.95         & 0.55$^{+0.35}_{-0.40}$/0.5 \\
log(n$_{e}$)            & 10.7$^{+0.43}_{-0.35}$/10.4         &  10.6$^{+0.46}_{-0.31}$/10.4        &  10.7$^{+0.38}_{-0.36}$/10.4        & 10.6$^{+0.48}_{-0.31}$/10.3 \\
$\tau_{H\alpha}$        & 0.05$^{+0.22}_{-0.04}$/0.05        &    0.05$^{+0.33}_{-0.04}$/0.03      & 0.05$^{+0.12}_{-0.04}$/0.08         & 0.03$^{+0.25}_{-0.02}$/0.02  \\
\hline
	\end{tabular}
\end{table*}
\noindent As the parameter histograms give no information on dependencies between the parameters, we investigate those in the following. We generate 2D histograms of all possible parameter pairs to infer the parameter dependencies. In Fig.~\ref{dep1} we show the dependencies for all parameters for spectrum 109. The parameter dependencies for spectra 110-112 are shown in the appendix (see Figs.~\ref{dep2}, ~\ref{dep3}, ~\ref{dep4}). For both, prominence and filament cases, we find:\\
\textit{a)} for heights $\geq$200000~km ($\geq$0.88~R$_{\ast}$), thickness shows maximum bins being $<$ 50000~km ($<$0.22~R$_{\ast}$); \\
\textit{b)} thickness increases with decreasing area, maximum bins are always found at thicknesses $<$ 50000~km ($<$0.22~R$_{\ast}$) and area $>$ 0.5 for spectrum 109, 110, and 111, for spectrum 112 we find maximum bins also at area $>$ 0.65;\\
\textit{c)} for heights of 500000~km (2.22~R$_{\ast}$), log(n$_{e}$) shows maximum bins of 10.3-10.4;\\
\textit{d)} for heights of 500000~km (2.22~R$_{\ast}$), temperature shows maximum bins in the range of 70000-100000~K; \\
\textit{e)} log(n$_{e}$) increases with decreasing temperature, maximum bins are found for temperatures in the range of 80000-110000~K and log(n$_{e}$) in the range of 10.4-10.5;\\
\textit{f)} for $\tau_{H_{\alpha}}$ between 0.01-0.1, thickness shows a maximum bin of $<$50000~km ($<$0.22~R$_{\ast}$);\\
\textit{g)} log(n$_{e}$) decreases with increasing thickness, maximum bins are found for thickness $<$ 100000~km ($<$0.44~R$_{\ast}$) and log(n$_{e}$) of 10.4-10.9;\\
\textit{h)} for heights of 500000~km (2.22~R$_{\ast}$), $\tau_{H_{\alpha}}$ shows maximum bins $<$1; \\
\textit{i)} $\tau_{H_{\alpha}}$ increases with decreasing temperature, maximum bins can be found for $\tau_{H_{\alpha}}<$1 and for temperatures in the range of 50000-75000~K; \\
\textit{j)} area increases with decreasing log(n$_{e}$), maximum bins are found for log(n$_{e}$) in the range of 10.3-10.4, and area ranging from 0.45-0.95~A$_{P}$/A$_{\ast}$;\\
\textit{k)} log(n$_{e}$) increases with increasing gas pressure, maximum bins are found at 1 (spectra 109, 110, 111) and at 2~dyn~cm$^{-2}$ (spectrum 112);\\
\textit{l)} temperature increases with increasing gas pressure, maximum bins are found for maximum values of both grids, i.e. at a temperature of 300000~K and a gas pressure of 5~dyn~cm$^{-2}$;\\
\textit{m)} thickness increases with decreasing gas pressure, maximum bins are found at low thickness ($<$50000~km corresponding to $<$0.22~R$_{\ast}$) and gas pressures of 3-5~dyn~cm$^{-2}$;\\
\textit{n)} $\tau_{H\alpha}$ increases with decreasing area.\\
From the 2D histograms one can see that the maximum bins, i.e. the bins with the largest number of cases, show values partly deviating from the median values and maximum bins from the 1D histograms, but nearly all are located within the percentiles of the median values (see section~\ref{discussion}).\\
As one can see from the histograms, we find solutions for both filament and prominence geometry. From table~\ref{paramvalues} one can see that we always find more matching cases for the filament geometry. Accordingly we can not exclude prominence geometry. Moreover, we find that the parameters deduced from modeling yield similar ranges for both geometries.\\
We model all four spectral signatures showing the largest line of sight velocities. The modeling results show that there are only marginal differences between the four event spectra, revealing no evolution of parameters such as e.g. area or height. Height can not be constrained because of a dominating thermal emission therm in the line source function (see section~\ref{solardiscuss}).

\section{Discussion}
\label{discussion}
\subsection{Parameter distributions}
\begin{figure}
\begin{center}
	\includegraphics[width=\columnwidth]{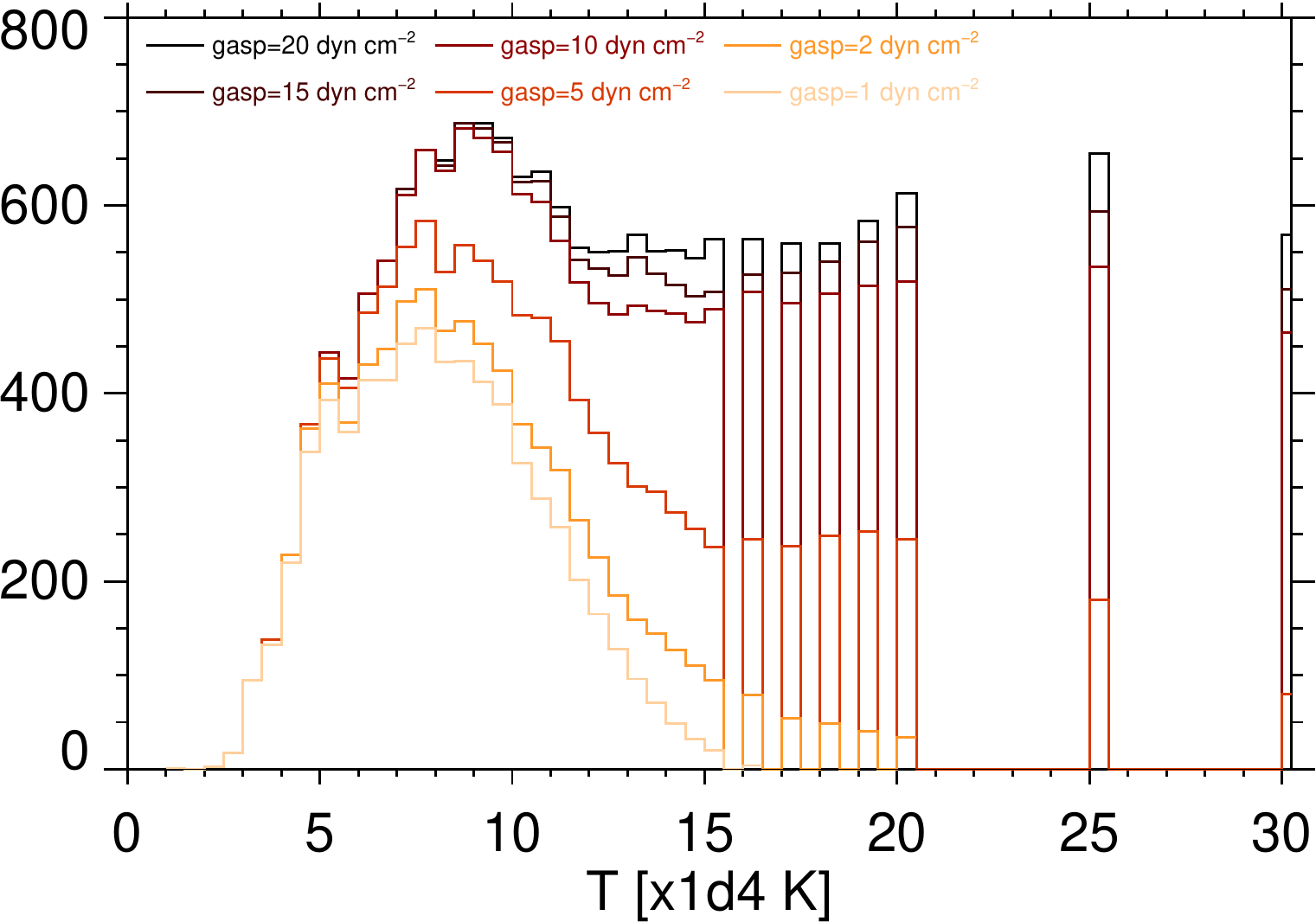}
    \caption{Dependence of gas pressure on temperature derived for spectrum 109 for the filament case.\label{gasvstemperature}}
\end{center} 
\end{figure}

Summarizing the results, we find from modeling that most cases matching the observations have median values of heights of 400000~km (1.77~R$_{\ast}$), thickness of 50000-75000~km (0.22-0.33~R$_{\ast}$), temperatures of 95000-115000~K, log(n$_{e}$) of 10.6-10.7, a gas pressure of 1-2~dyn~cm$^{-2}$, $\tau_{H\alpha}$ of 0.03-0.1, and an area of 0.55-0.7~A$_{\ast}$. The spread of the distributions, given in percentiles (10 and 90\%), show large values, as the distributions are broad. Furthermore, we also see that the maxima of the 1D distributions deviate from the median values (see table~\ref{paramvalues}), which is expected as the distributions are asymmetric. \\
In section~\ref{modelling} we already saw how the parameters are related among each other, we also noted that the maximum bins of the 2D histograms partly deviate from the median values in the 1D histograms. If we look into that in more detail we identify parameter combinations where the maximum bins lie at the edge of the percentiles.\\
Starting with the parameter height, the median is 400000~km (1.77~R$_{\ast}$) and the maximum bin is 500000~km (2.22~R$_{\ast}$). For nearly all parameter combinations we see that the maximum and neighbouring maximum bins, i.e. bins within 15\% of the maximum bin, are in the range of 400000-500000~km (1.77-2.22~R$_{\ast}$), i.e. still within the percentiles. The only case where the deviation is larger is for the height and thickness 2D histogram, here we see maximum bins down to 200000~km (0.88~R$_{\ast}$), which is also the first bin in the 1D histogram where the distribution significantly rises. But also this deviation is within the percentiles of the distribution.\\
For the parameter thickness, the median ranges from 50000-75000~km (0.22-0.33~R$_{\ast}$) and the maximum bin is $<$25000~km ($<$0.11~R$_{\ast}$), although the distribution is rather flat and the maximum bin not well pronounced. Here the maximum bins of the 2D distributions are $<$100000~km ($<$0.44~R$_{\ast}$) for all parameter combinations, 
all being well within the percentiles.\\
The median of log(n$_{e}$) ranges from 10.6-10.7 and the maximum bins from 10.3-10.4. Here, maximum bins of log(n$_{e}$) vs. thickness lie at the upper edge of the percentiles of the distribution. As the 1D distribution of log(n$_{e}$) is not broad, also the range of the percentiles is smaller, therefore even small deviations lie at the edges of the percentiles.\\
The median of temperature ranges from 95000-115000~K and the maximum bins from 75000-100000~K. All maximum bins lie within the percentiles except the distribution of temperature and thickness, where the temperatures are at or even slightly below the lower value of the percentiles. For the distribution of temperature and gas pressure we find also maximum bins at 300000~K, so way beyond the percentiles (see below).\\
The median of gas pressure ranges from 1-2~dyn~cm$^{-2}$ and the maximum bins from 2-3~dyn~cm$^{-2}$. Also here we find cases (gas pressure vs. thickness, gas pressure vs. temperature) where maximum bins lie at the edges of the percentiles.\\
The median of area ranges from 0.55-0.7~A$_{\ast}$ and the maximum bins from 0.5-1.0~A$_{\ast}$. Here all cases show maximum bins lying at the upper edges of the percentiles (except for spectrum no.~112). For the distribution of area and gas pressure there are also maximum bins at the lower edges of the percentiles. As the distribution of area is broad cases are found at both upper and lower edges of the grid.\\
The median of $\tau_{H\alpha}$ ranges from 0.03-0.1 and the maximum bins from 0.02-0.16. All cases show maximum bins lying within the percentiles, except for the distribution of $\tau_{H\alpha}$ and gas pressure, where the maximum bins lie at or slightly below the lower edges of the percentiles.\\
We see that maximum bins from parameter combinations including gas pressure lead to large ranges of maximum bins of the second parameters. This behaviour is especially evident for parameter combinations temperature vs. gas pressure and area vs. gas pressure. Taking a closer look at the dependence of gas pressure on temperature reveals that the lower the gas pressure the lower is the maximum of the temperature distribution. In Fig.~\ref{gasvstemperature} we plot the temperature distributions derived from different levels of extended gas pressure grids, with maximum values of 1 (light-red) to 20 (black), from which this behaviour can be seen very well. What is also obvious from Fig.~\ref{gasvstemperature} is that the temperature maximum of the distribution shifts to lower temperatures for lower gas pressures. All of the maximum bins in the 2D histograms we see at large temperatures (250000, 300000~K) are caused by simultaneously large gas pressures. Therefore it is necessary to constrain at least one of these parameters independently to a reasonable physical range, because of their interdependence, which enables solutions even with unrealistically high values.

\begin{figure}
\begin{center}
	\includegraphics[width=\columnwidth]{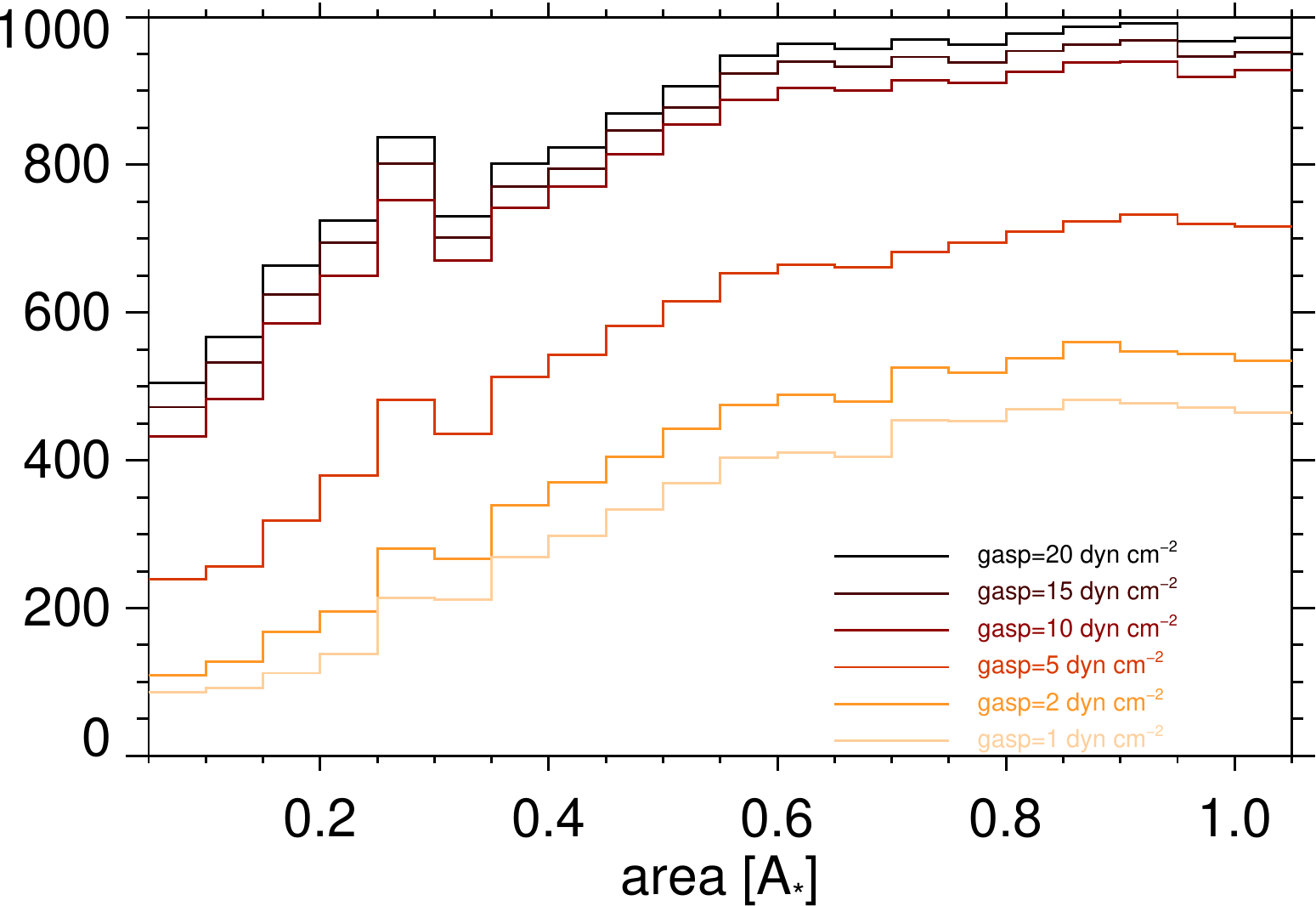}
    \caption{Dependence of gas pressure on area derived for spectrum 109 for the filament case.\label{areavstemperature}}
\end{center} 
\end{figure}
Even if the 2D histogram of area vs. gas pressure shows a broad range of maximum cases, gas pressure has no relationship with area in the same manner as for temperature and gas pressure. Increasing values of gas pressure have no influence on the distribution of area except that the total number of matching cases is also increased (see Fig.~\ref{areavstemperature}).

\subsection{How ``solar-like'' is the eruption?}
\label{solardiscuss}
From the results and discussion of parameter distributions we derive the following possible scenario:\\
The ejected material was observed at an already larger height with a median of 400000~km. A travelled distance of 400000km corresponds to roughly two stellar radii of V374~Peg. From the median value of thickness (50000-75000~km) we find that the median of height is a factor of 5-8 larger. The area has a median of 0.55-0.7~A$_{\ast}$. From a geometrical point of view this is a reasonable scenario, because one would expect that the material also expands while travelling through the stellar astrosphere, similar as on the Sun \citep{Maricic2009}. This may also explain the large FWHM of the emission features showing up to $\sim$200~km~s$^{-1}$. This width can not be explained by Doppler broadening alone which is $\sim$60~km~s$^{-1}$ (for a temperature of 100000~K and a micro-turbulence velocity of 5~km~s$^{-1}$). This geometrical configuration found from modeling combines small $\tau_{H\alpha}$ being $<$ 0.1, a temperature being much larger (95000-115000K) than known from quiescent filaments on the Sun (10000~K, see below), electron densities being reasonable, as such ranges (log(n$_{e}$)=10.6-10.7) are also observed on the Sun, and gas pressure (1-2~dyn~cm$^{-2}$) being just slightly higher than for typical solar cases (up to 1~dyn~cm$^{-2}$).\\
From investigations of solar quiescent filaments/prominences we know that those are cool structures with temperatures of 5000-10000~K \citep{Parenti2014}. They are located in the lower corona and have projected areas of $<$0.1\% of the solar disk \citep{Hao2015}. Their heights are typically $\sim$ 26000~km \citep{Wang2010} but can range between 10$^{4}$ and 10$^{5}$~km \citep{Labrosse2010}. Their thickness can vary from few hundreds up to few ten thousands of km \citep{Heinzel1996, Schwartz2004}. Their electron densities have been determined to be in the range of log(n$_{e}$)=9-11 \citep{Parenti2014}. Gas pressures are in the range of 0.02-1~dyn~cm$^{-2}$ \citep{Labrosse2010}.\\
If we compare these parameter ranges to the results derived from cloud modeling applied to the stellar case we identify the following (we do not comment on height as we compare quiescent solar prominence parameters to a stellar eruptive event) for \\
\textit{thickness:} the derived thickness from cloud modeling shows a median of 75000~km which reflects the upper range of solar filament/prominence thicknesses (few hundreds up to few ten thousands of km) \\
\textit{electron density:} here also the electron densities derived from modeling (median(log(n$_{e}$))=10.6-10.7) reflect the upper range of solar filament/prominence electron densities (log(n$_{e}$)=9-11).  \\
\textit{gas pressure:} here also the derived gas pressures from modeling (median=1) reflect the upper range of solar filament/prominence gas pressures (0.02-1~dyn~cm$^{-2}$).  \\
\textit{area:} this parameter is hard to compare as it is expected that the moving structure is expanding, however, the area of the stellar feature must be significantly larger than on the Sun to be detected in stellar spatially integrated observations. Solar filaments occupy $<$0.1\% of the solar disk, whereas the results from cloud modeling yield a median area of 55-70\% of the disk of V374~Peg. Investigations of stellar prominences have revealed areas of up to 20-30\% of the disk area \citep[e.g.][]{CollierCameron1990, Dunstone2006b, Leitzinger2016} for non-eruptive prominences. V374 Peg is a star able to form slingshot prominences, i.e. prominences located up to several stellar radii from the star \citep{Villarreal2018}, and it is predicted that they would also be observable \citep{Waugh2021}, although they have not been detected up to now on this star. 
 If a star's Alfvén radius exceeds its co-rotation radius, it is capable of hosting material in the corona, i.e. slingshot prominences \citep{Villarreal2018}. The ejection mechanism of slingshot prominences may be different from  the solar one driven by magnetic reconnection. Such prominences would be centrifugally ejected when their mass, which is continuously supplied by the stellar wind, becomes too large to be confined by the magnetic field \citep{Jardine2019}. So far, slingshot prominence disappearance has been reported, i.e. when signatures of slingshot prominences did not appear anymore after several rotation periods of repeated detection \citep[e.g.][]{Dunstone2006a}, but up to now has not been observed directly. 
A slingshot prominence scenario may explain that we derive already for the first event spectrum (spectrum no.109) a median height of 400000~km, i.e. that the eruption of plasma started already from large heights. Additionally, the prominence moves a distance of at least 120000~km per spectrum (projected bulk velocity of $\sim$400~km~s$^{-1}$ and exposure time of 300~s), i.e. 480000~km during all four event spectra, which is consistent with the determined large heights.\\
\textit{temperature:} here the median value for the stellar case ranges from 95000-115000~K which is way above the 10000~K for quiescent solar prominences. But here we need to consider that we compare quiescent solar prominence temperatures with temperatures of a stellar moving cloud. What we know from the Sun is that when a filament erupts it gets heated and partly ionized while travelling through the corona. \citet{Heinzel2016} present the analysis of a hot CME core at a height of 2.2~R$_{\sun}$, i.e. an erupting prominence, having a temperature, as derived from NLTE modeling, of $\sim$100000~K, the same as derived for the stellar case here. \citet{Heinzel2016} derived these temperatures from modeling observations of the CME core from UV spectral lines. On the Sun the CME core can be only observed off disk, because of the bright solar disk. \\
\subsection{Filaments versus prominences}
As shown from our results we can not distinguish between filament and prominence geometry for this event. This is surprising as from the Sun we know that filaments only appear in absorption in the Balmer lines whereas prominences appear only in emission. On the Sun, the line source function of filaments/prominences is dominated by scattering of the incident solar radiation. The line source function generally consists of the scattering term which strongly depends on the incident radiation, and the thermal emission term (spontaneous emission after collisional excitation). Compared to the Sun, the incident radiation of V374~Peg at a given height above its surface is lower at Balmer line wavelengths, as for dM stars the maximum of emission shifts to longer wavelengths, due to their lower temperature and their intrinsic faintness. For the stellar case, the second term of the source function is not negligible and seems to even dominate. Otherwise the emission features in the V374~Peg spectra could only be explained for prominence geometry and not for filament geometry. But here we show that both geometries are possible, which indicates that the thermal emission term dominates over the scattering term for both geometries and therefore enables a filament case seen in emission. This second term of the line source function is dependent on electron density and temperature \citep{Heinzel2015}. Electron density is within the range solar values, but the temperature is significantly higher and the incident radiation level is low because of the large height and different spectral type of the star compared to the Sun. The prominence/filament height can therefore not be determined reliably for this event as the dilution factor (being a factor of height) is part of the scattering term only, which is negligible relative to the thermal emission term, i.e. it has only limited effect. Moreover the incident radiation of V374~Peg also only affects the scattering term and therefore uncertainties in the stellar spectrum can not have a significant effect on the results, because of the negligible contribution.\\
Summarizing, the blue-shifted emission features found by \citet{Vida2016} can be interpreted in terms of a hot eruptive filament/prominence with other parameters being in the upper range of solar prominence parameters.\\

\section{Conclusions}
We investigate the blue-shifted extra emissions found in Balmer line spectra of the dMe star V374~Peg presented in \citet{Vida2016}. We apply a cloud model for both filament and prominence case, and we determine the Balmer line source function from 1D NLTE modeling. We derive 1D and 2D histograms revealing the parameter distributions, as well as the dependencies of the parameters among each other. We find that all parameters, except temperature and area, lie at the upper part of the parameter ranges of solar quiescent prominences/filaments. We conclude therefore that the most likely scenario is a hot eruptive stellar filament (or prominence) on V374~Peg. The matching modeling solutions for the filament case are more numerous than for the prominence case but this does not set any preference on the type of geometry. NLTE modeling further revealed that emission features on the blue side of Balmer line spectra in M dwarfs can be indeed caused by erupting filaments, which are per definition in front of the stellar disk, and appear only in absorption on the Sun.\\ 



\section*{Acknowledgements}
M.L. and P.O. acknowledge the Austrian Science Fund (FWF): P30949-N36 for supporting this project. P.H., M.L. and P.O acknowledge support from the Czech Science Foundation, grant
19-17102S. P.H. was supported by the program ``Excellence Initiative - Research University'' for years 2020-2026 at University of Wroclaw, project No.
BPIDUB.4610.96.2021.KG.



\section*{Data Availability}
For the present study we have used data from the Polarbase archive which is publicly available at \url{http://polarbase.irap.omp.eu/}.
 



\bibliographystyle{mnras}
\bibliography{Mybibfile} 




\appendix

\section{Appendix}




\newpage
\begin{figure*}
\begin{center}
	\includegraphics[width=8cm]{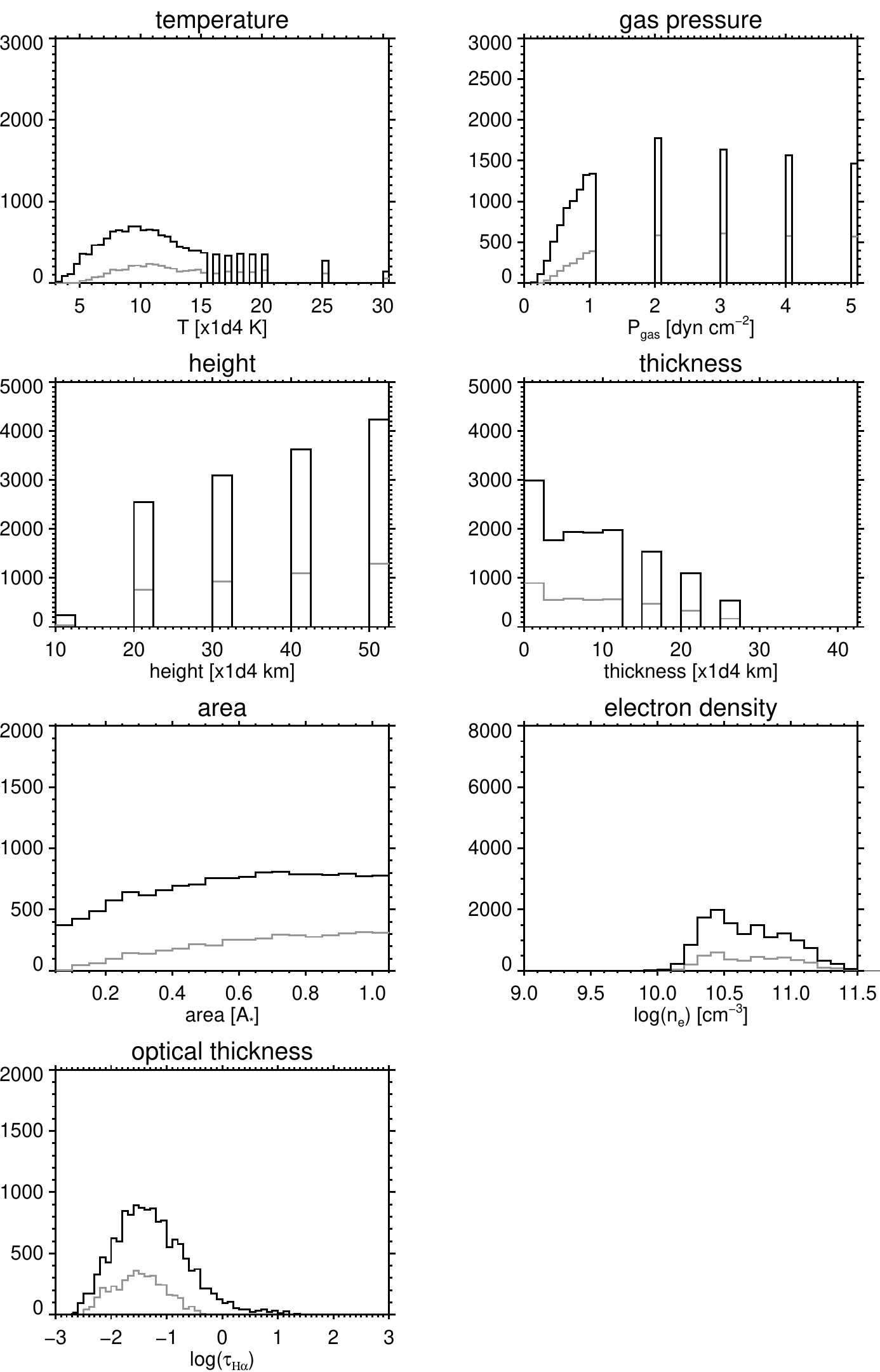}
	\includegraphics[width=8cm]{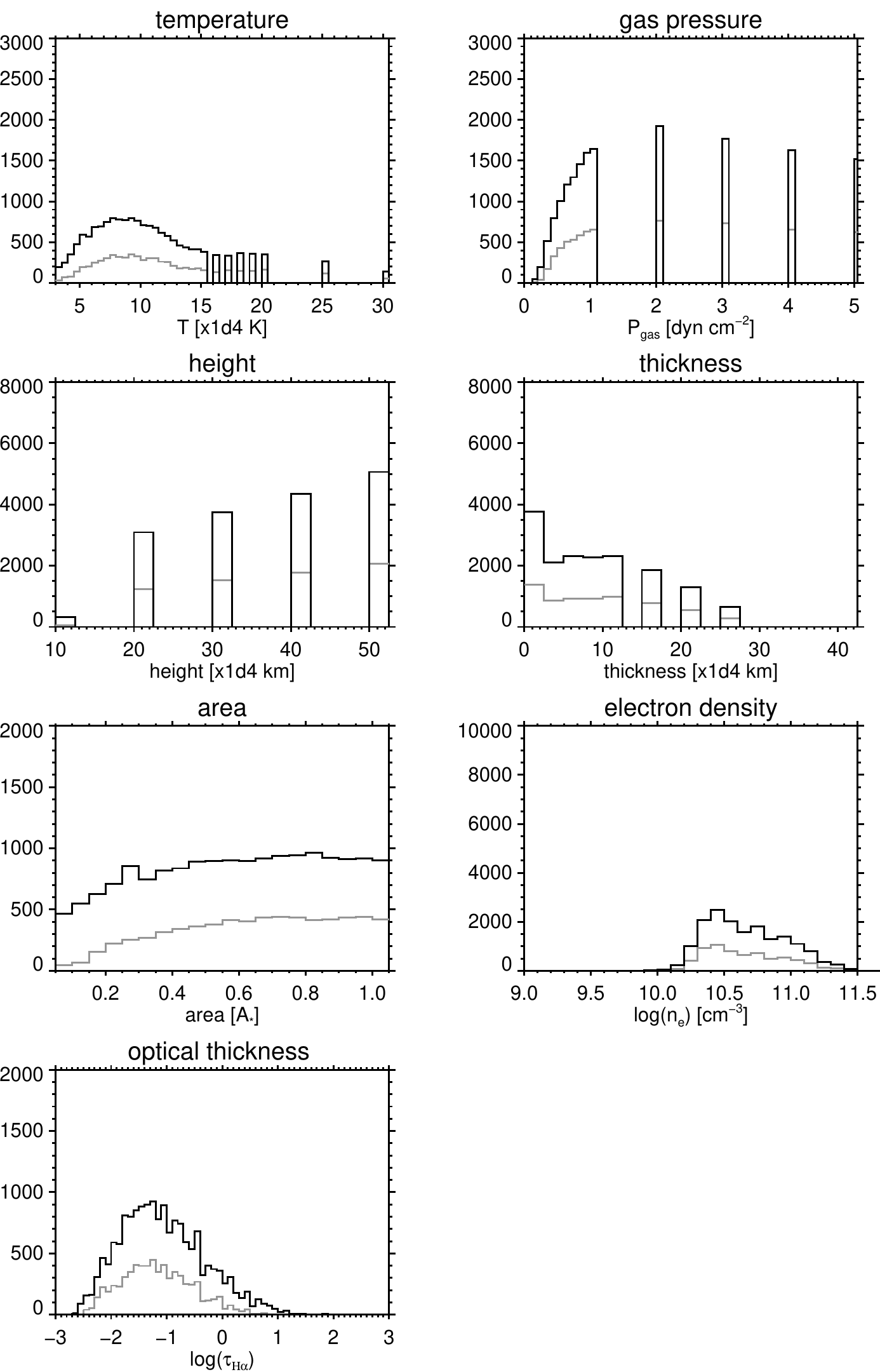}
    \caption{Histograms of the parameters of cloud model results for spectrum 110. Columns 1 and 2: Histograms of the prominence case. Columns 3 and 4: Histograms of the filament case. Black solid lines refer to 2-$\sigma$ results whereas grey solid lines refer to 1-$\sigma$ results. \label{histograms1d239}}
\end{center} 
\end{figure*}

\newpage
\begin{figure*}
\begin{center}
	\includegraphics[width=8cm]{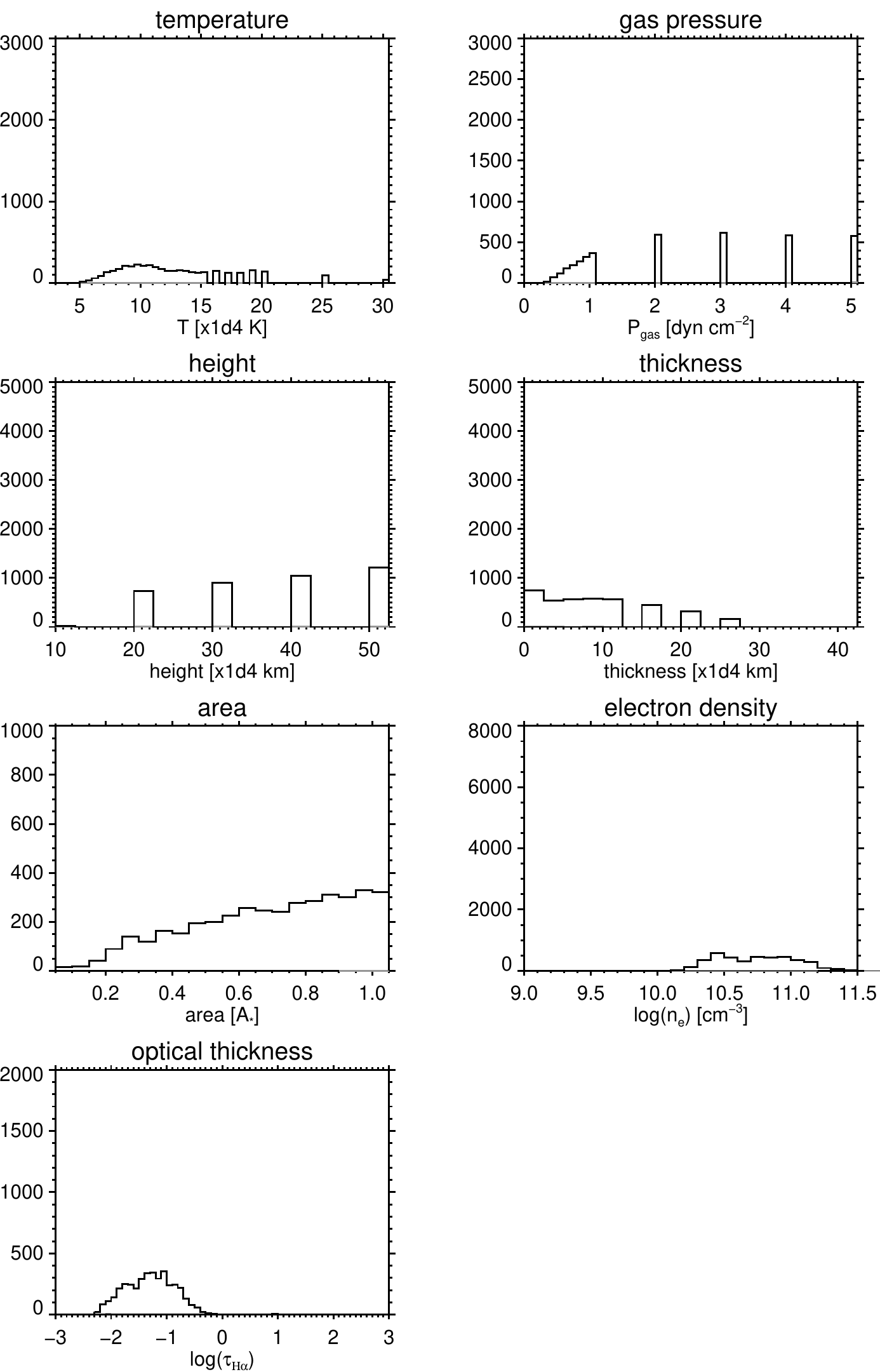}
	\includegraphics[width=8cm]{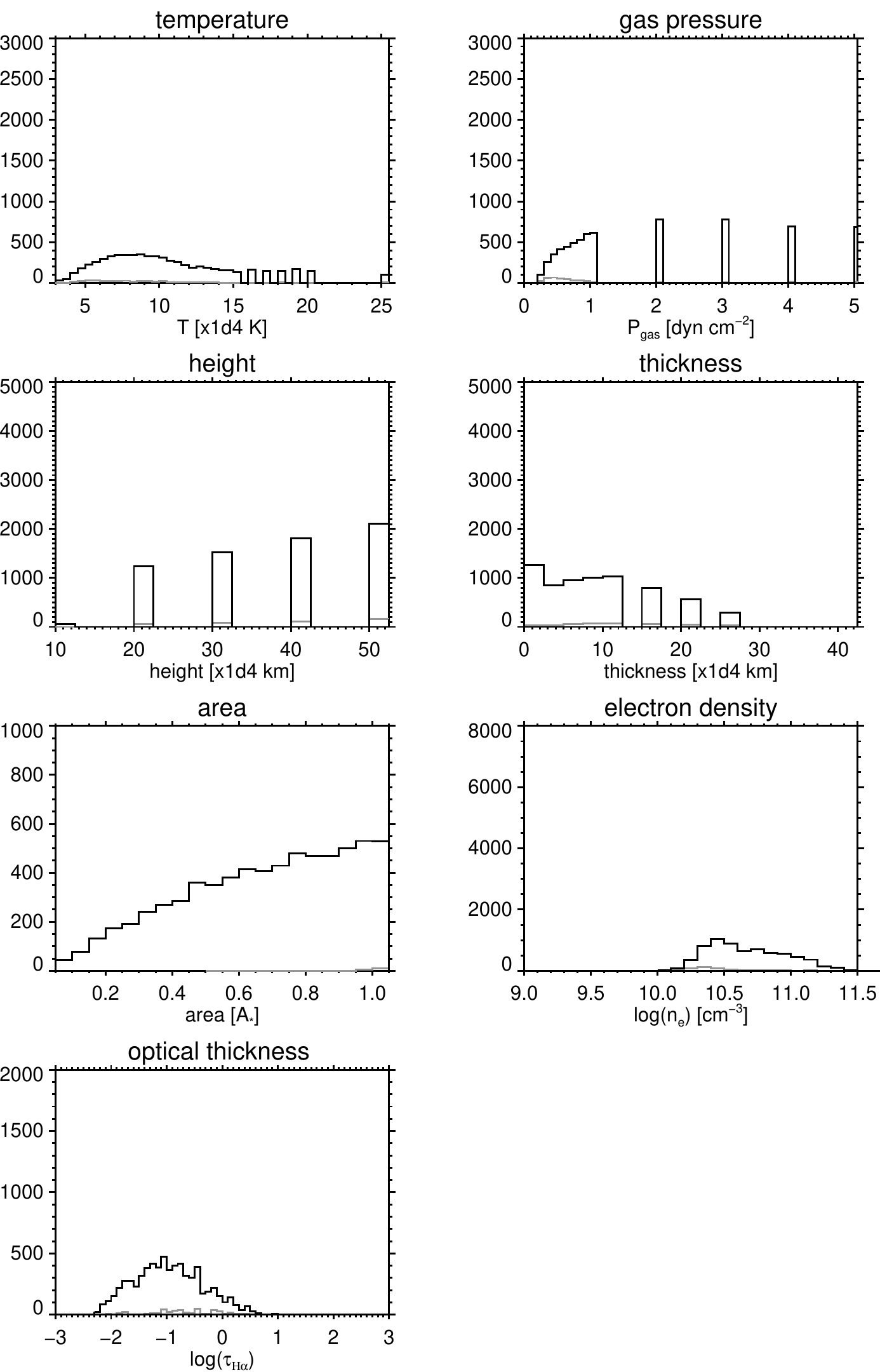}
    \caption{Histograms of the parameters of cloud model results for spectrum 111. Columns 1 and 2: Histograms of the prominence case. Columns 3 and 4: Histograms of the filament case. Black solid lines refer to 2-$\sigma$ results whereas grey solid lines refer to 1-$\sigma$ results. \label{histograms1d240}}
\end{center} 
\end{figure*}

\newpage
\begin{figure*}
\begin{center}
	\includegraphics[width=8cm]{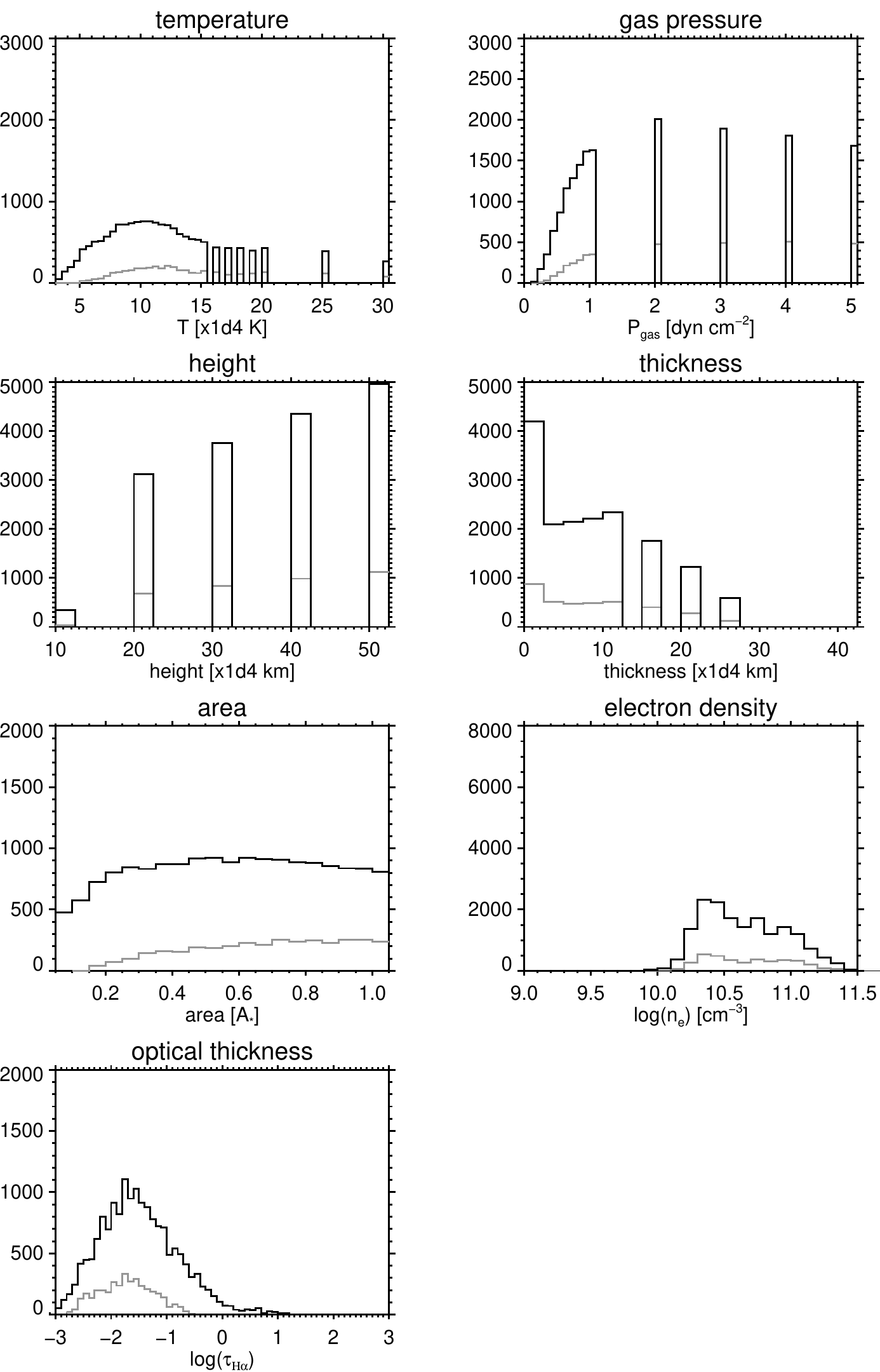}
	\includegraphics[width=8cm]{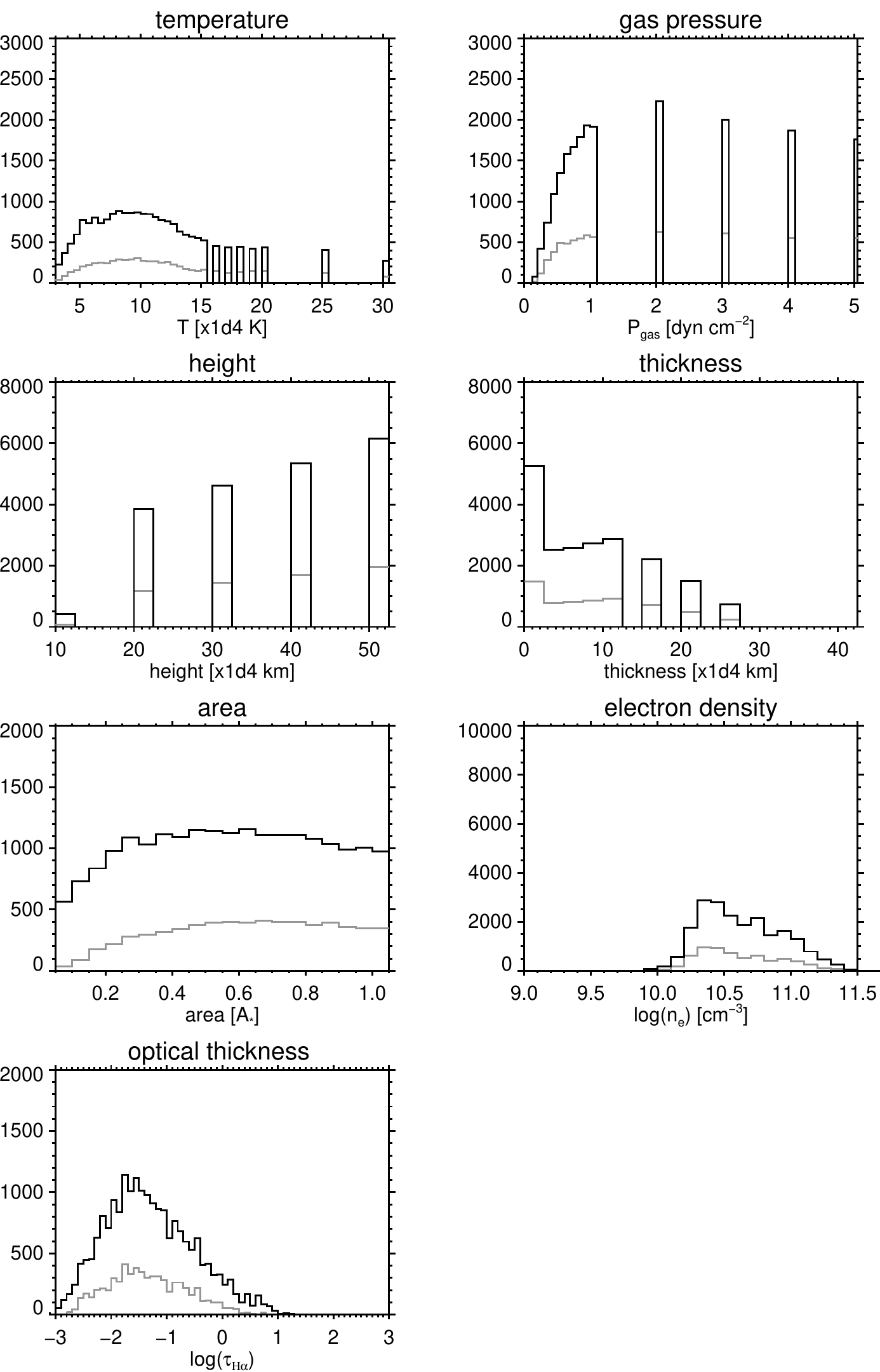}
    \caption{Histograms of the parameters of cloud model results for spectrum 112. Columns 1 and 2: Histograms of the prominence case. Columns 3 and 4: Histograms of the filament case. Black solid lines refer to 2-$\sigma$ results whereas grey solid lines refer to 1-$\sigma$ results. \label{histograms1d241}
}
\end{center} 
\end{figure*}


\newpage
\begin{figure*}
\begin{center}
	\includegraphics[width=7cm]{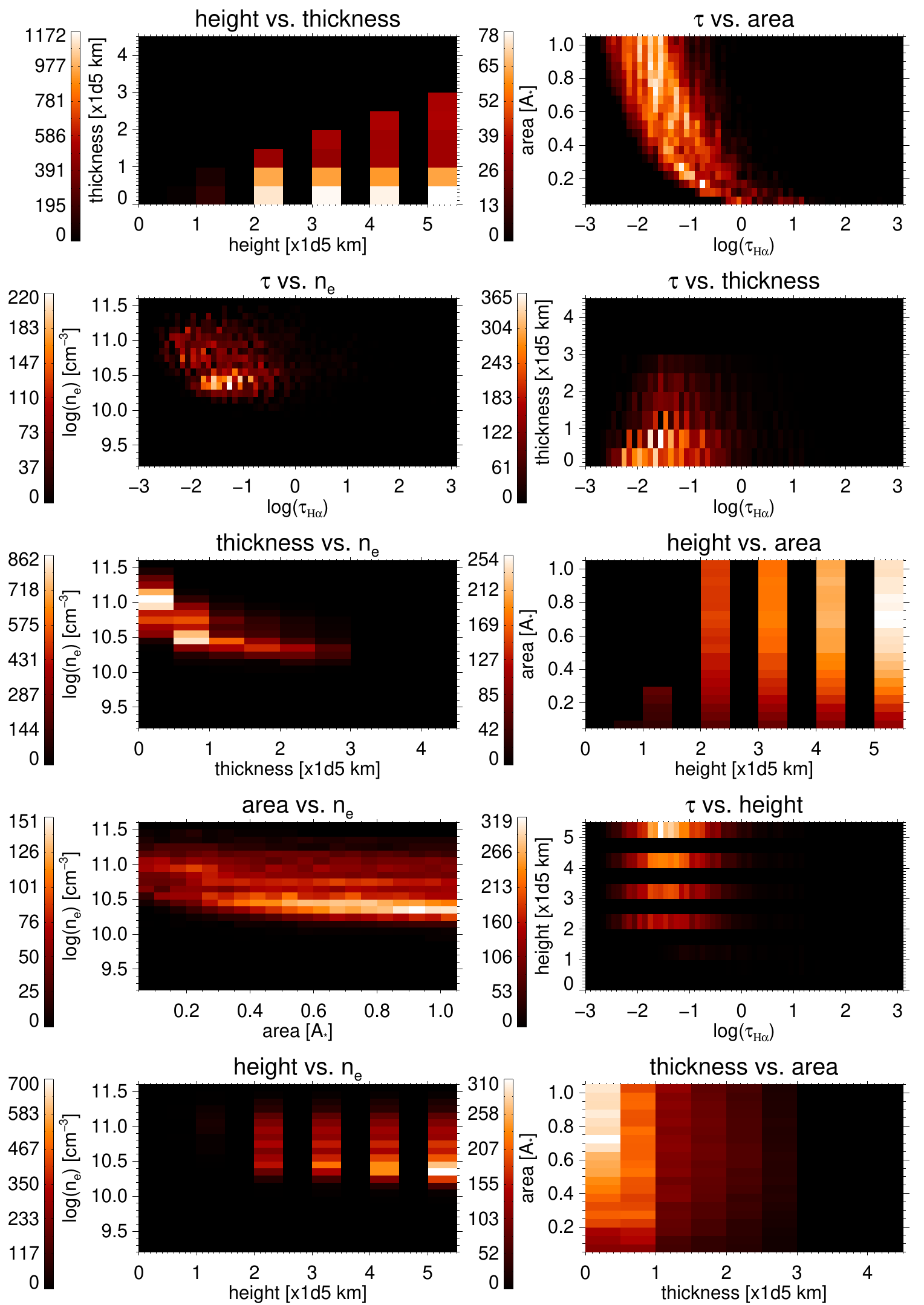}
	\includegraphics[width=7cm]{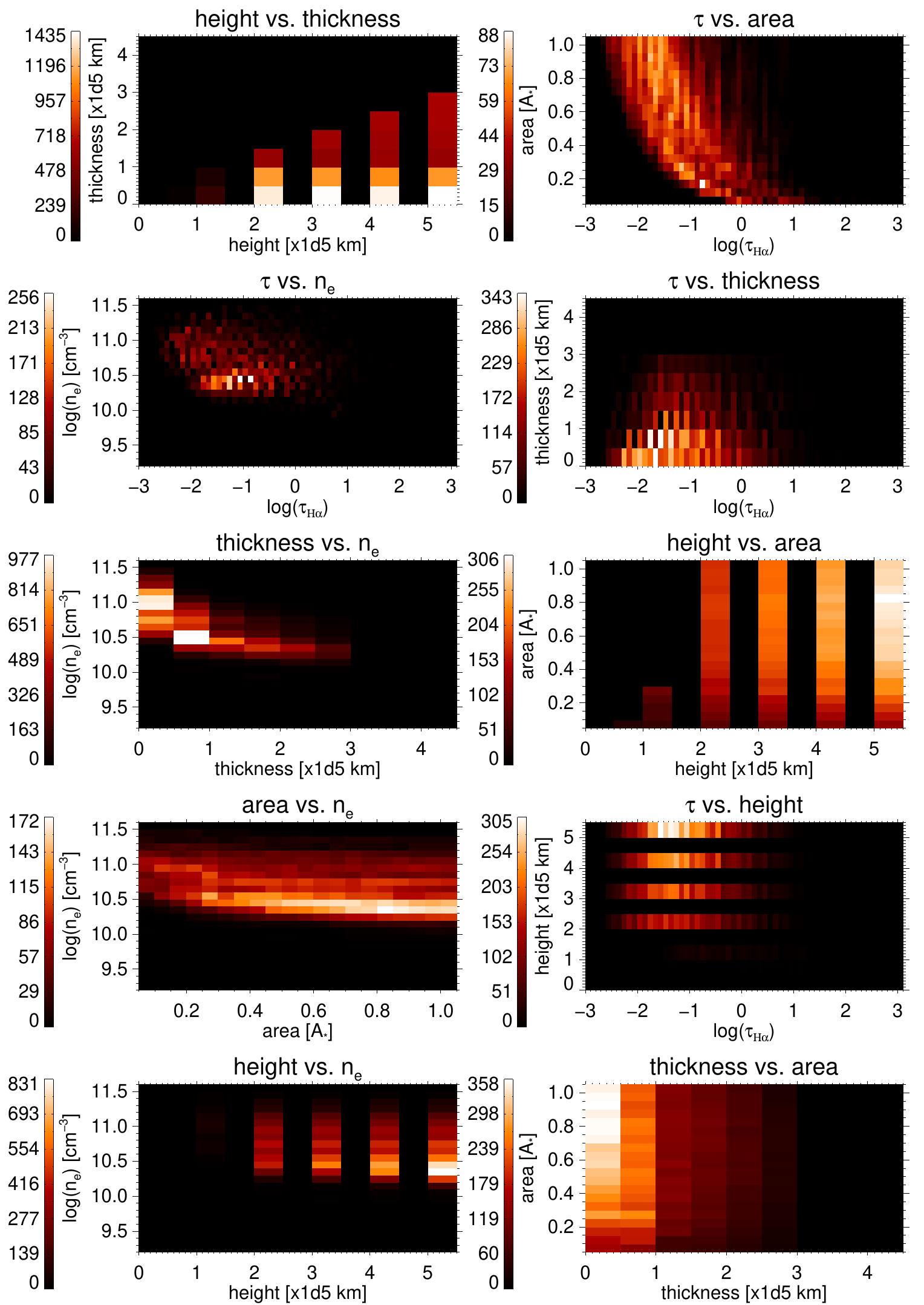}
	\includegraphics[width=7cm]{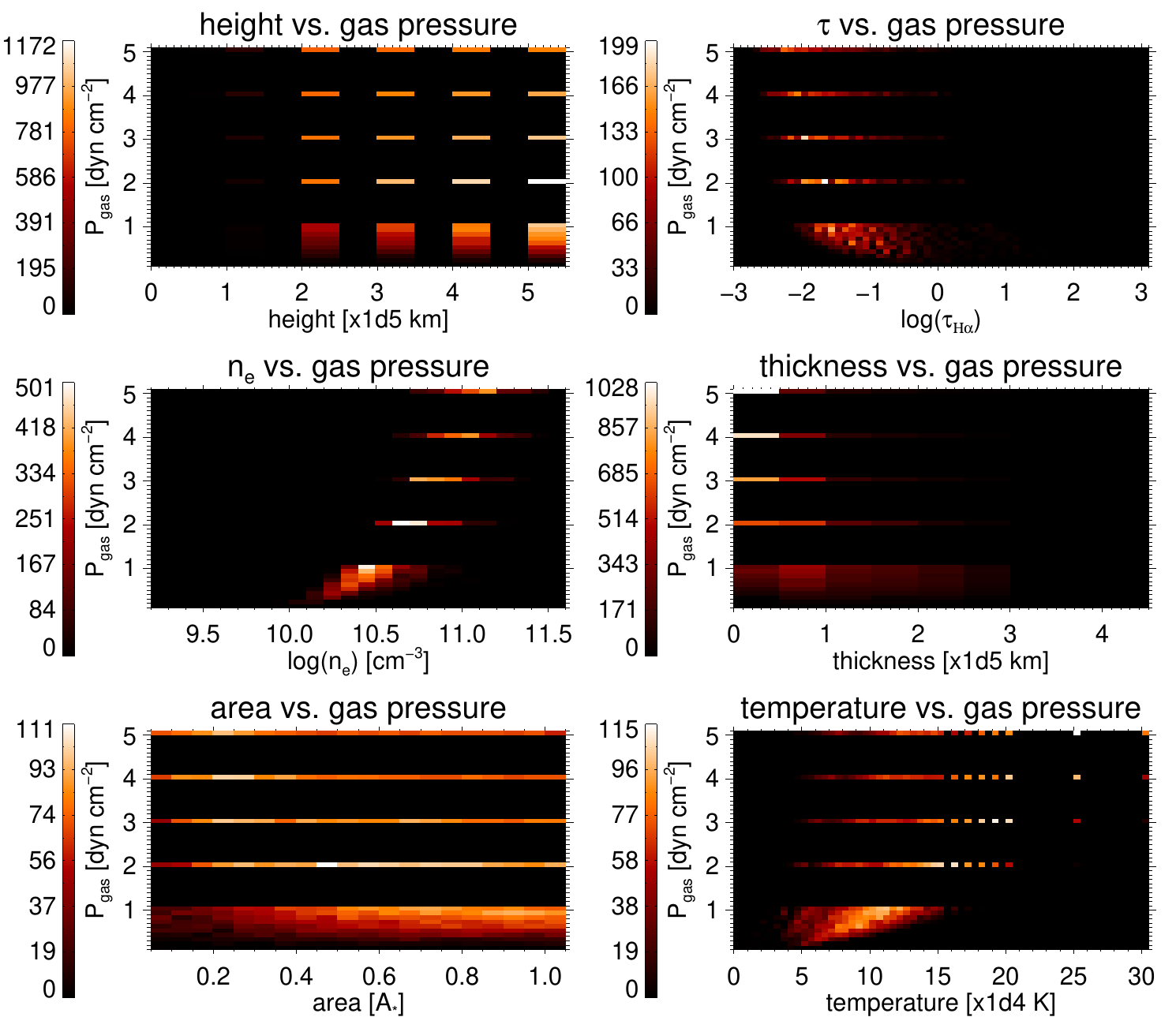}
	\includegraphics[width=7cm]{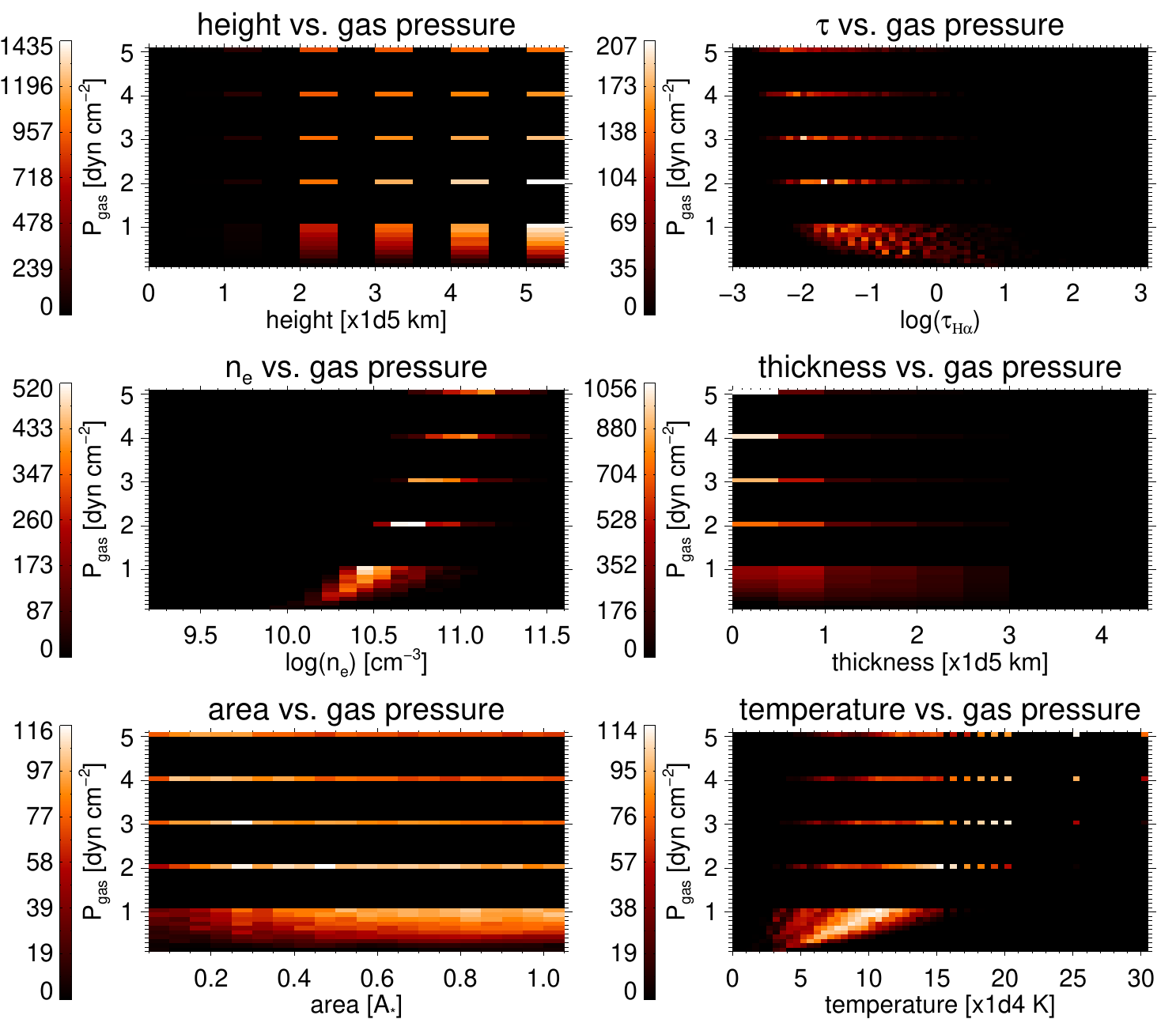}
	\includegraphics[width=7cm]{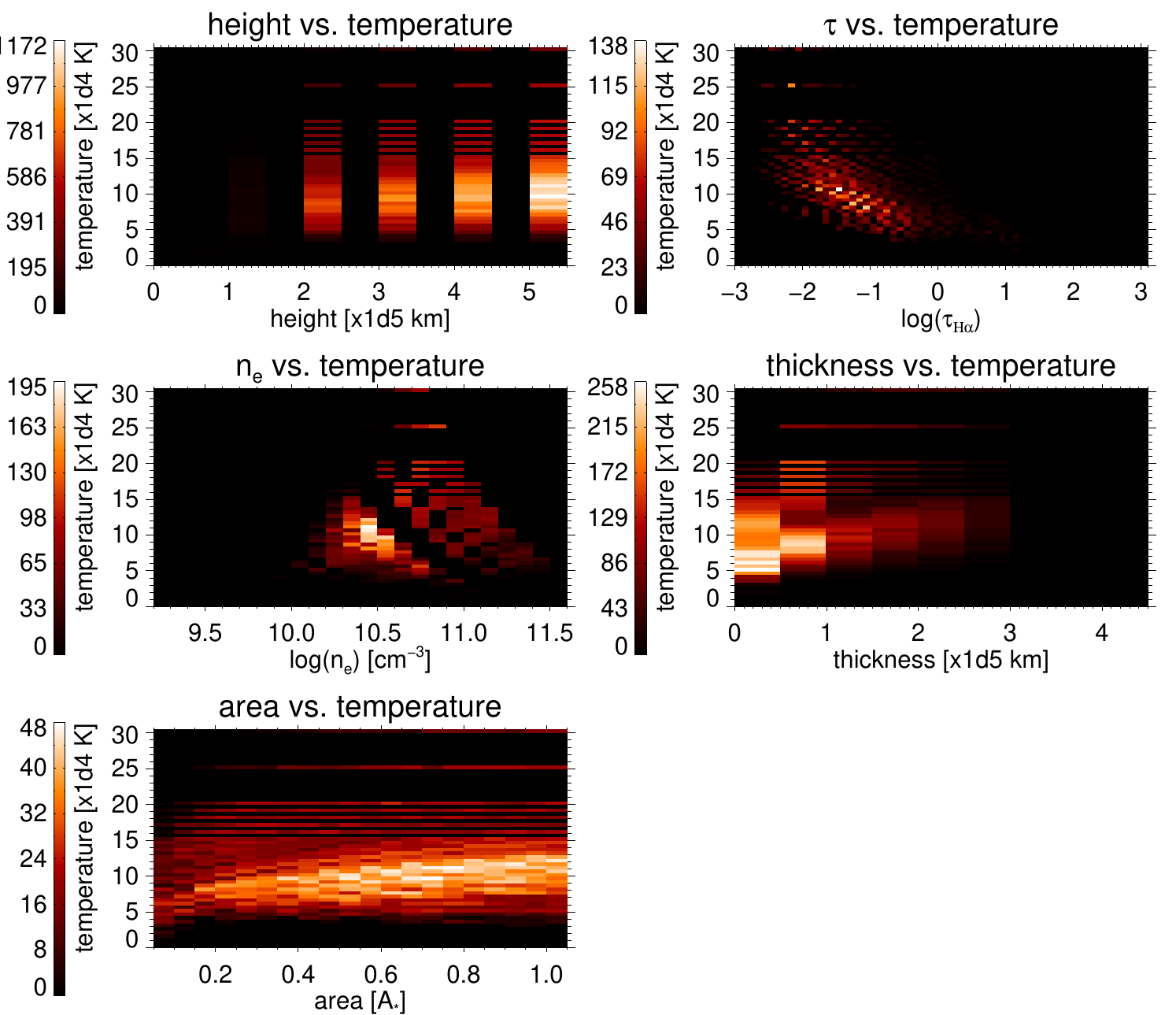}
	\includegraphics[width=7cm]{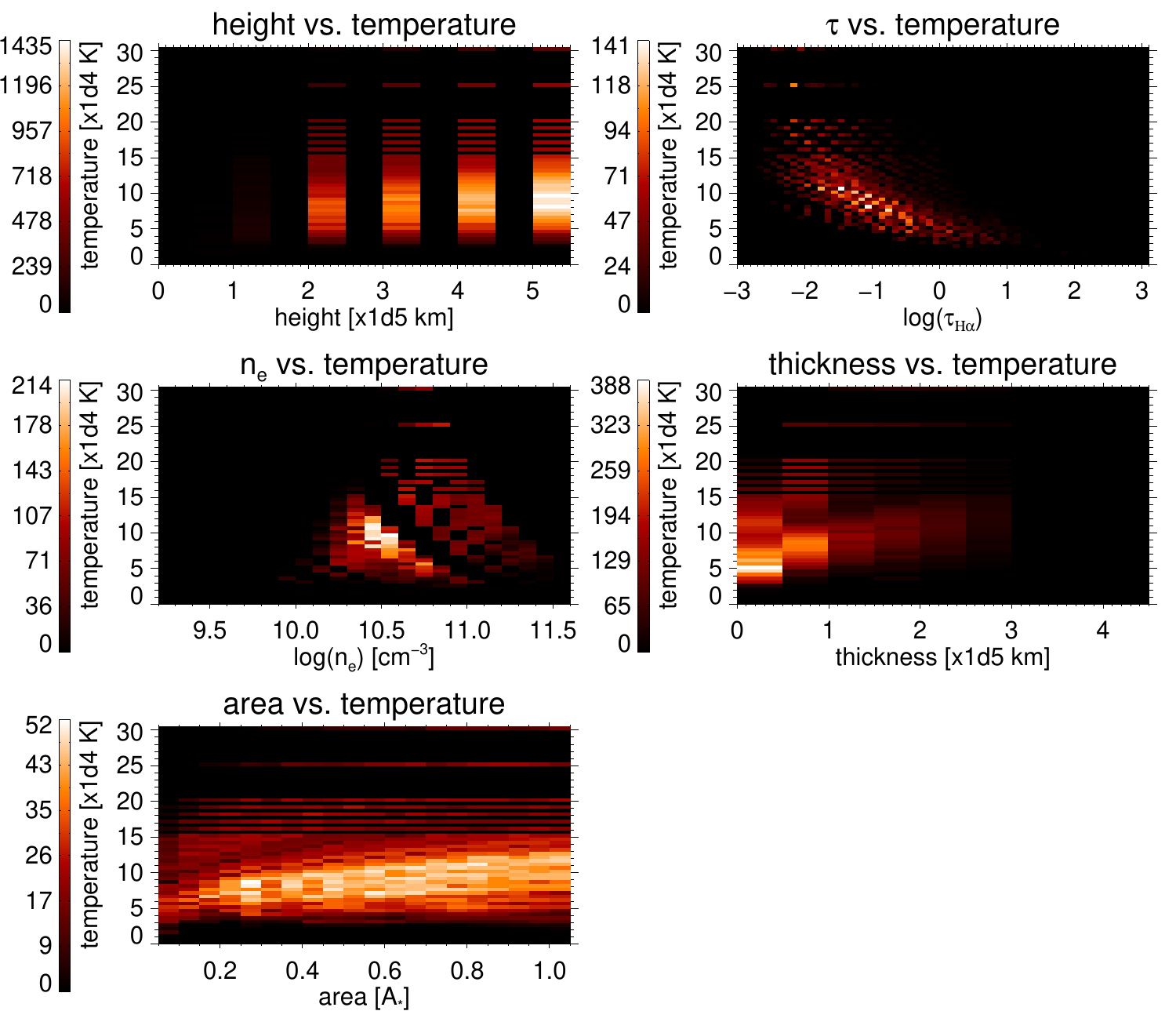}
    \caption{2D histograms of all parameter combinations of the 2-$\sigma$ cloud model results for spectrum no.~110. Columns 1 and 2: 2D histograms of all parameter combinations of the 2-$\sigma$ cloud model results for the prominence case. Columns 3 and 4: 2D histograms of all parameter combinations of the 2-$\sigma$ cloud model results for the filament case. \label{dep2}}
\end{center}
\end{figure*}

\newpage
\begin{figure*}
\begin{center}
	\includegraphics[width=7cm]{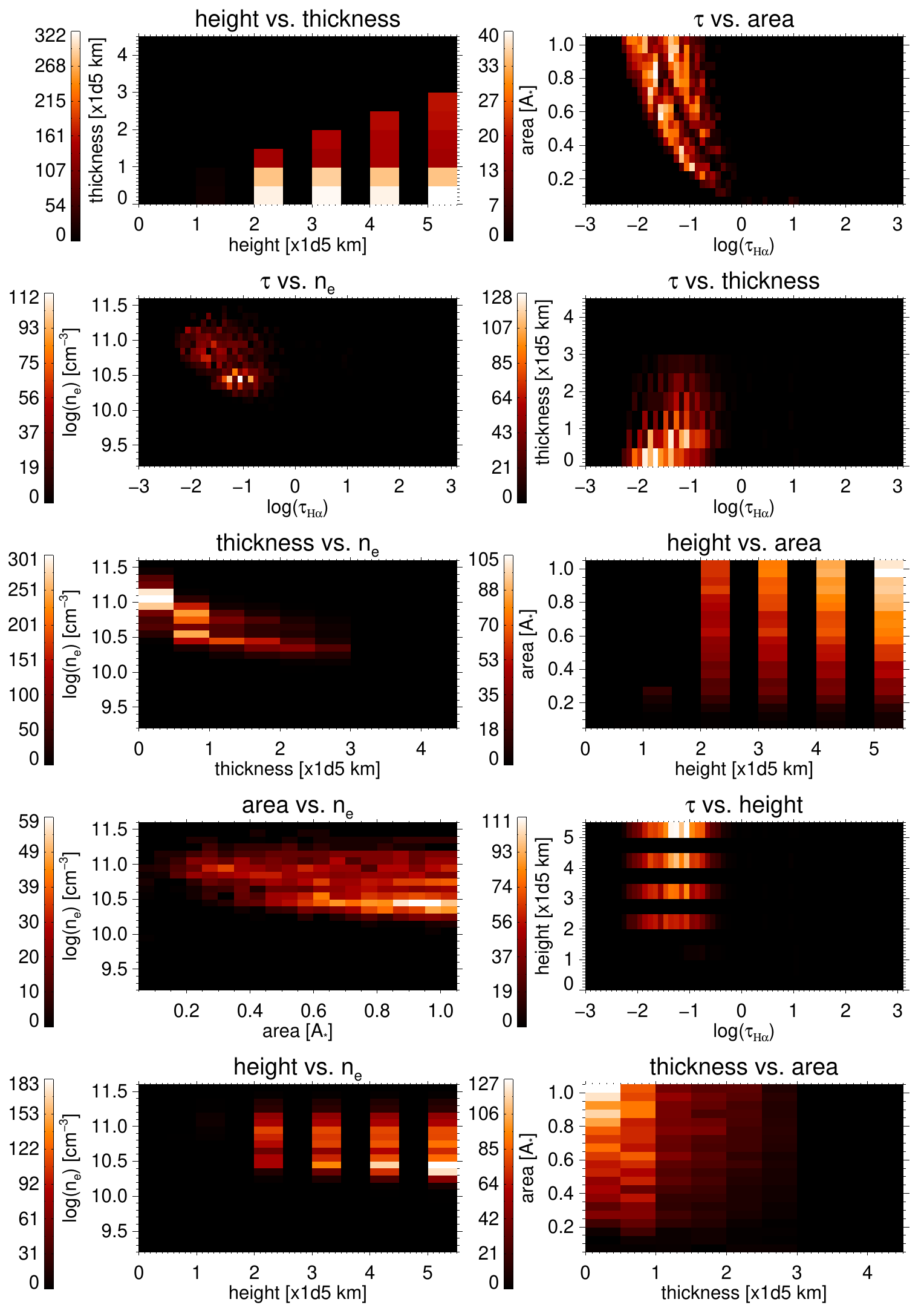}
	\includegraphics[width=7cm]{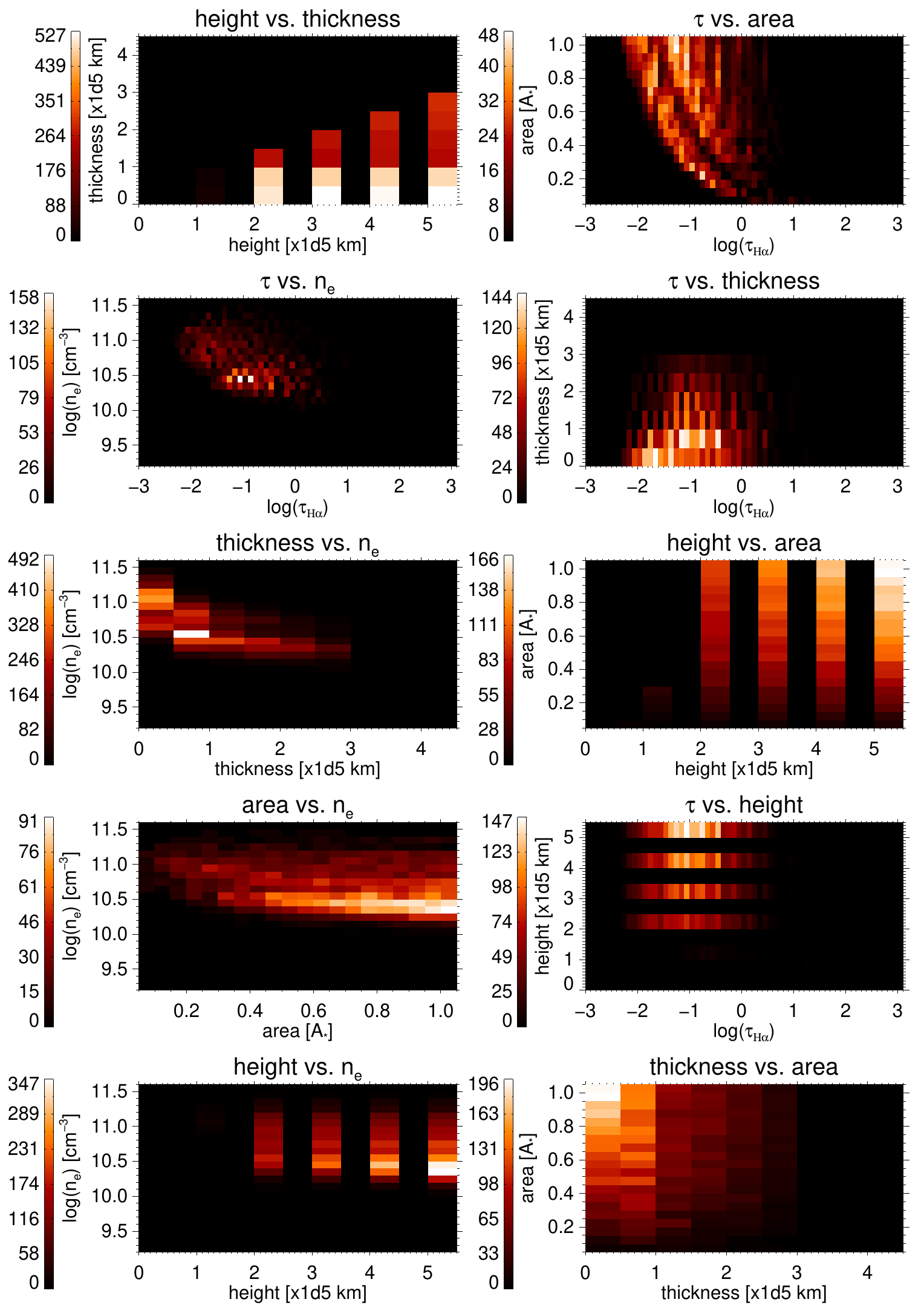}
	\includegraphics[width=7cm]{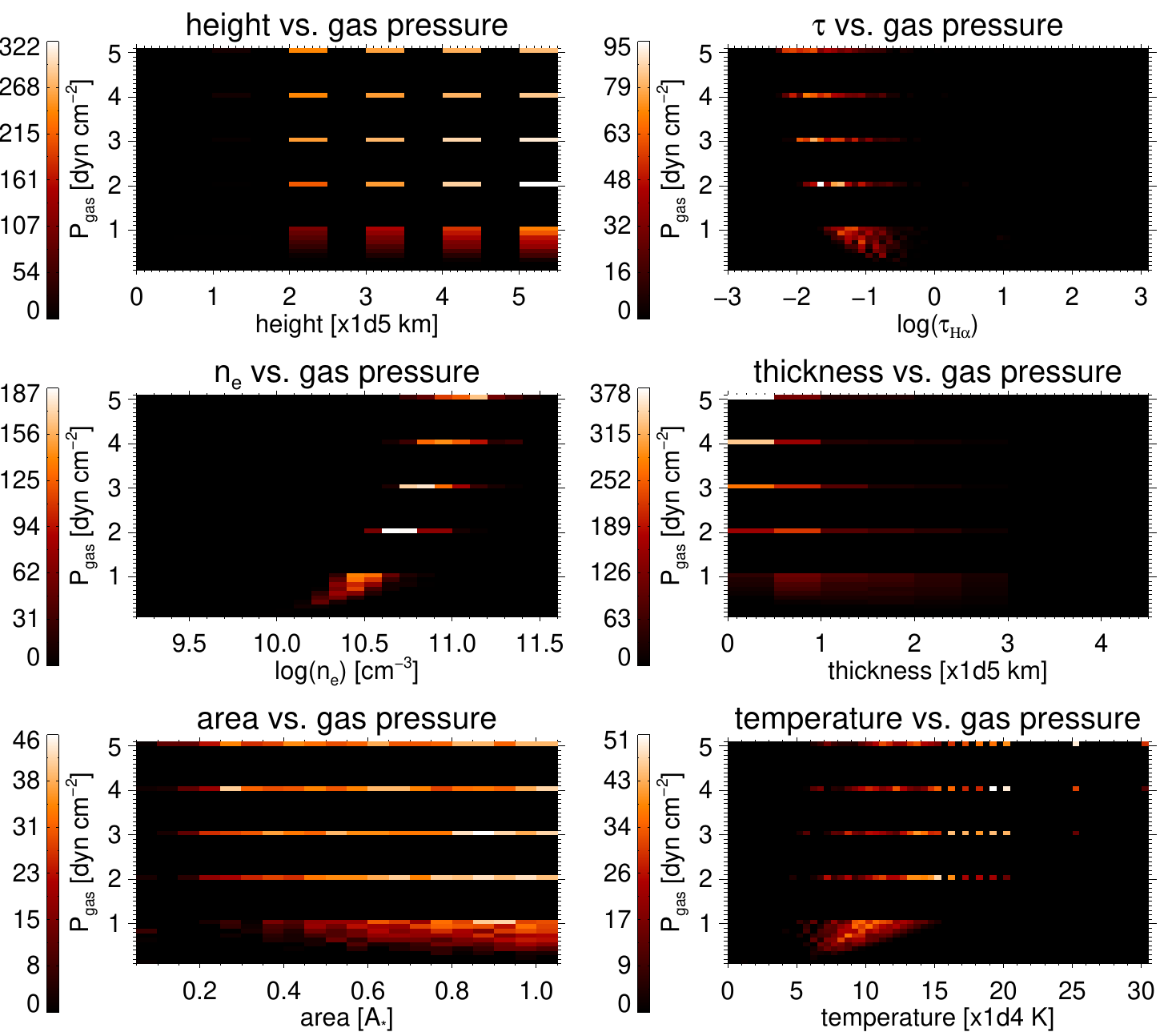}
	\includegraphics[width=7cm]{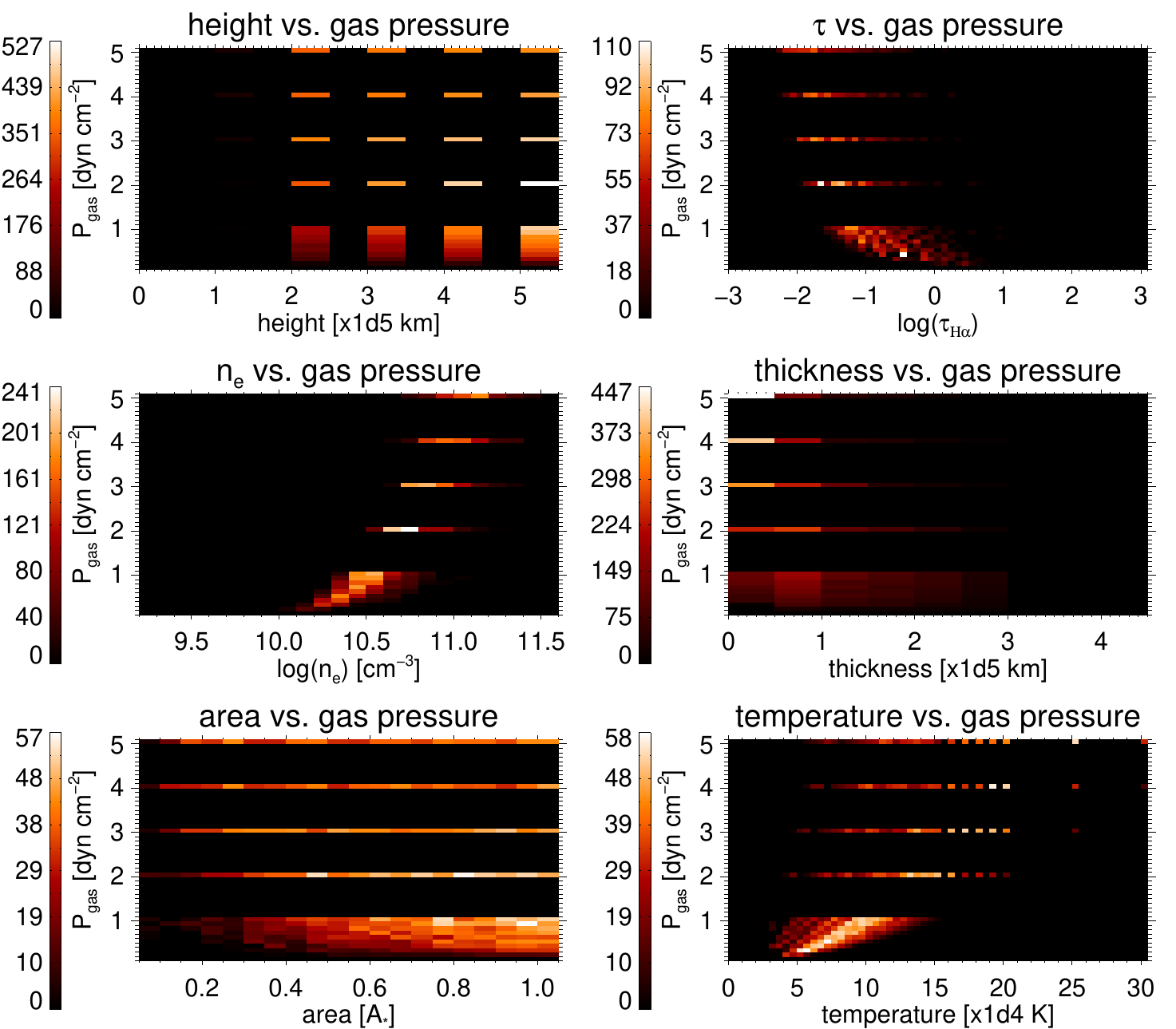}
	\includegraphics[width=7cm]{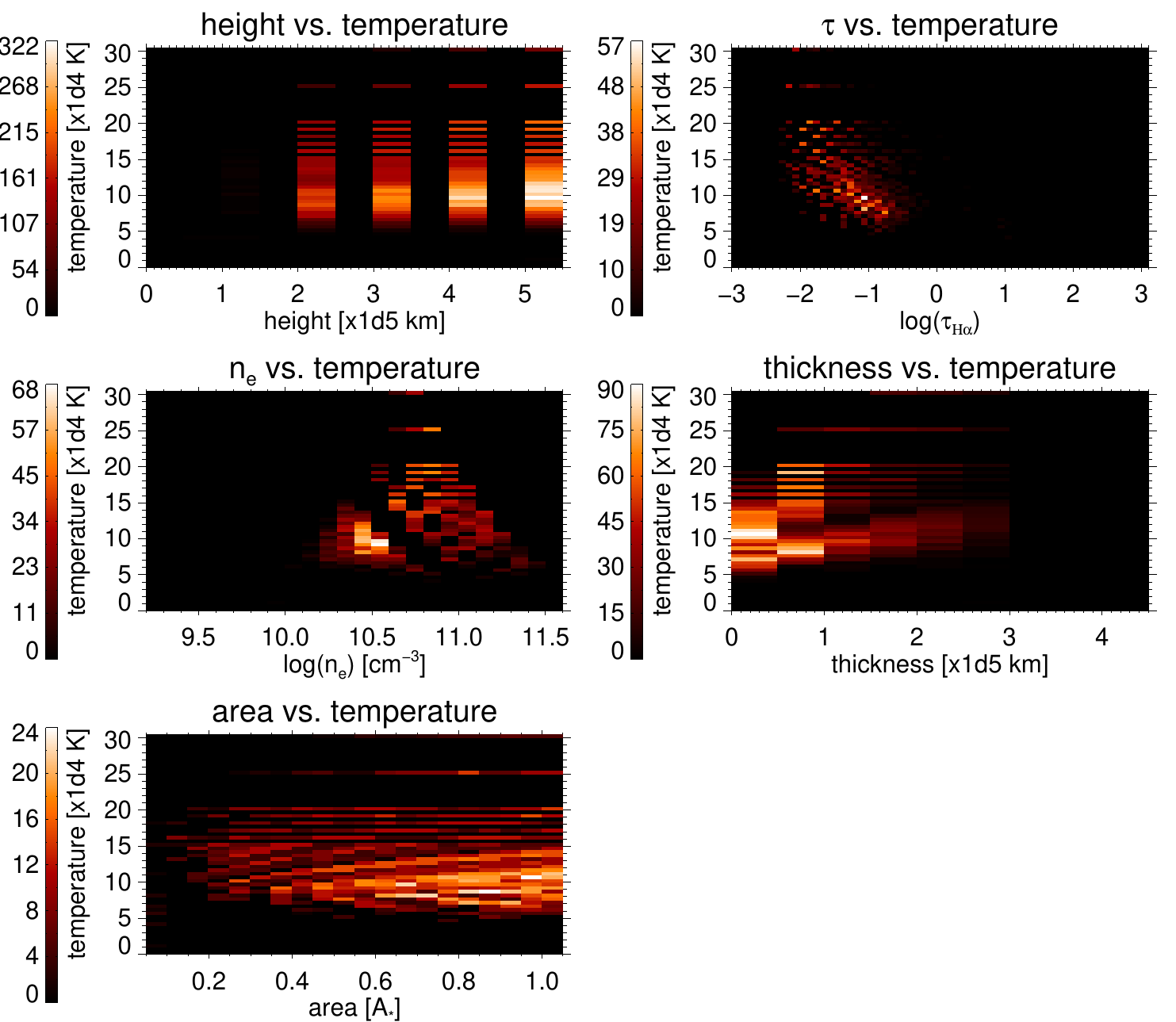}
	\includegraphics[width=7cm]{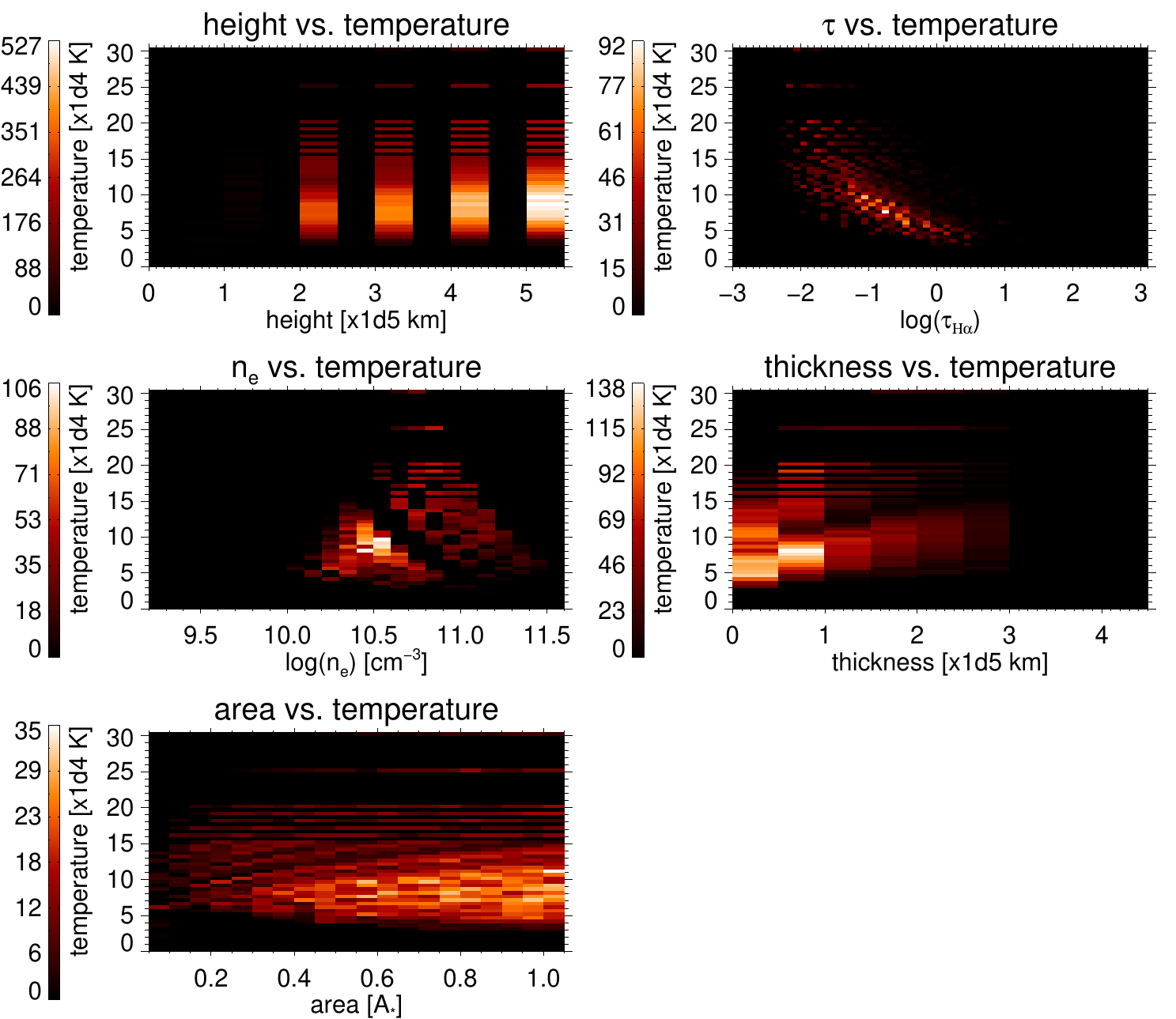}
    \caption{2D histograms of all parameter combinations of the 2-$\sigma$ cloud model results for spectrum no.~111. Columns 1 and 2: 2D histograms of all parameter combinations of the 2-$\sigma$ cloud model results for the prominence case. Columns 3 and 4: 2D histograms of all parameter combinations of the 2-$\sigma$ cloud model results for the filament case. \label{dep3}}
\end{center}
\end{figure*}


\newpage
\begin{figure*}
\begin{center}
	\includegraphics[width=7cm]{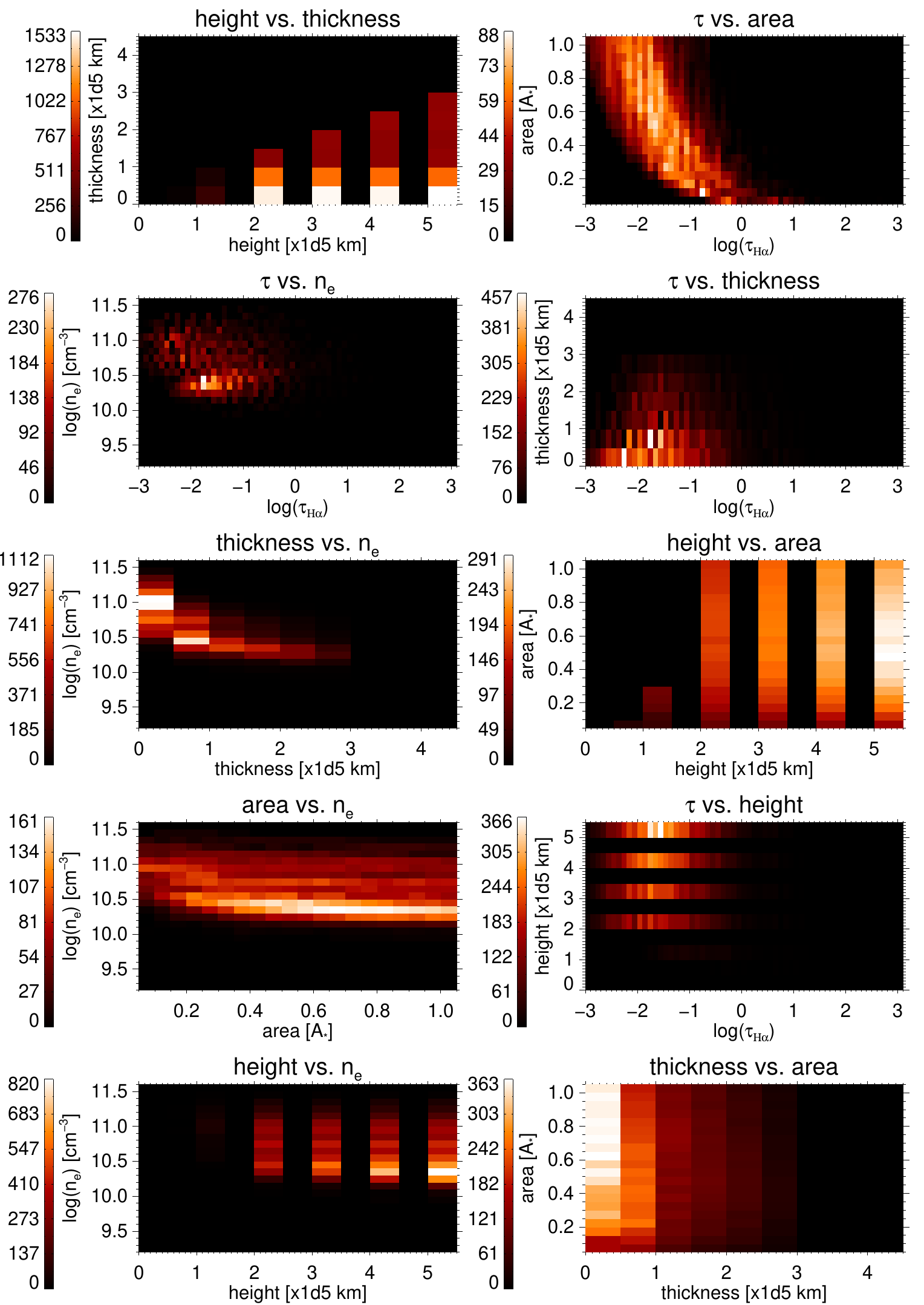}
	\includegraphics[width=7cm]{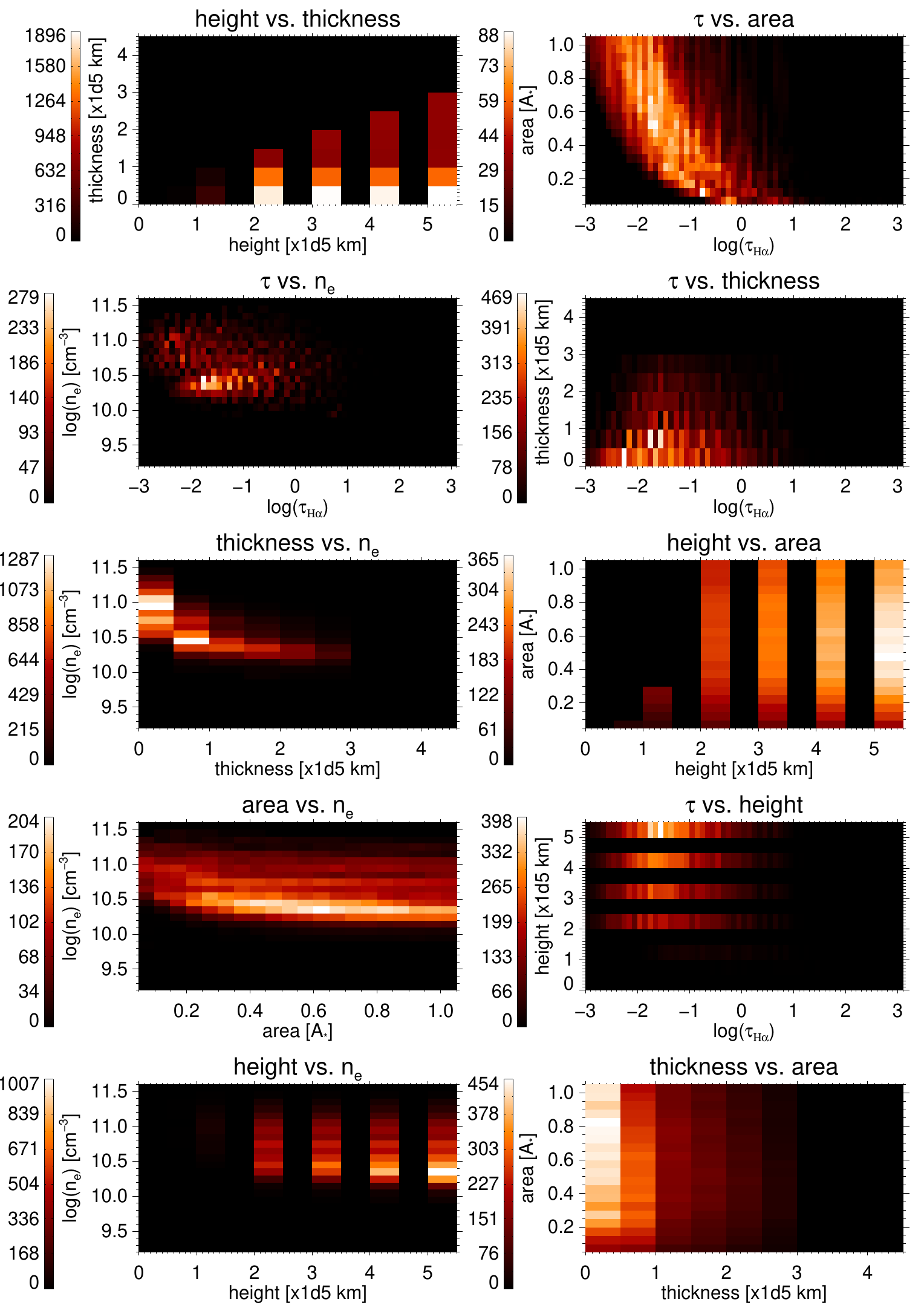}
	\includegraphics[width=7cm]{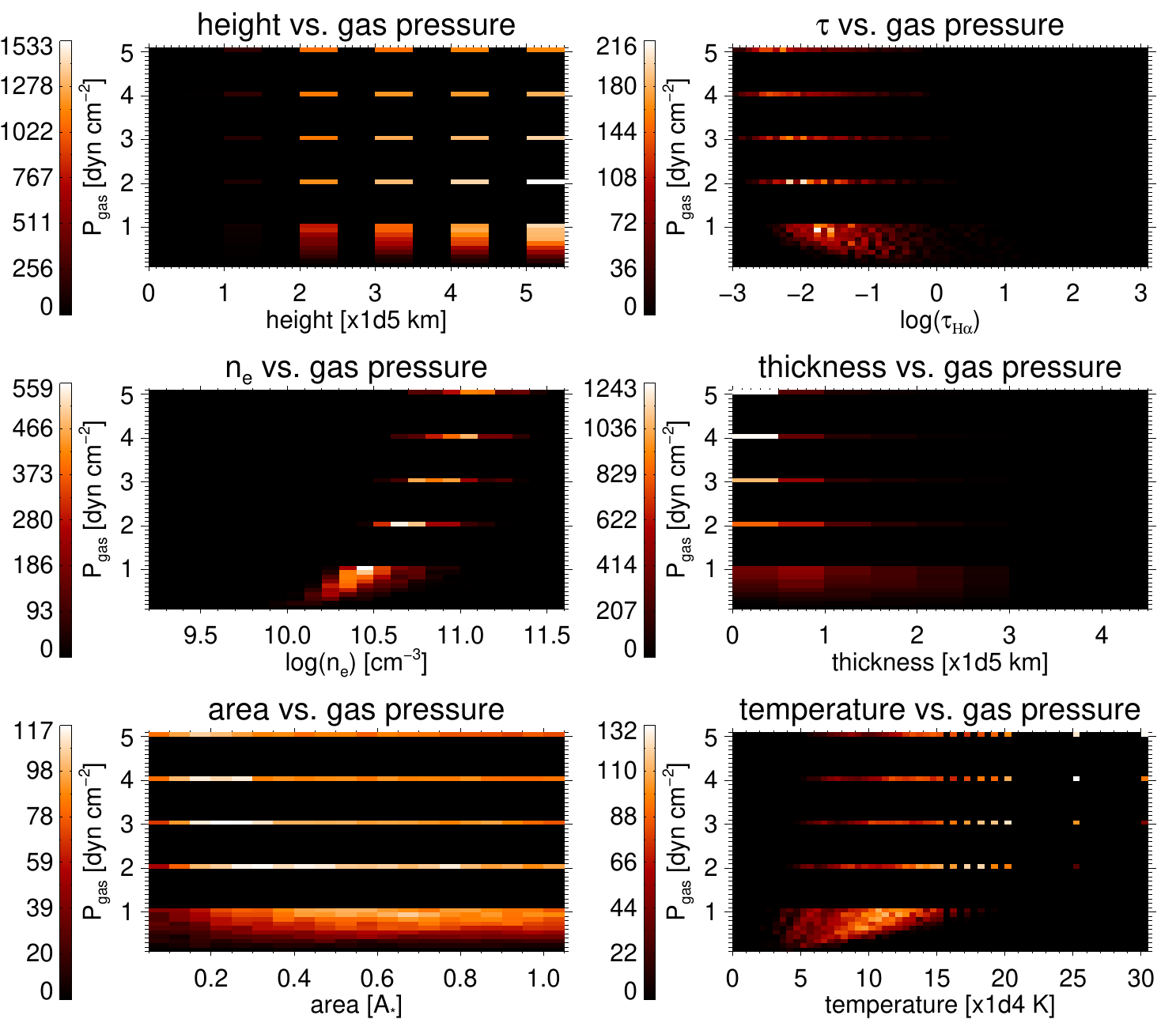}
	\includegraphics[width=7cm]{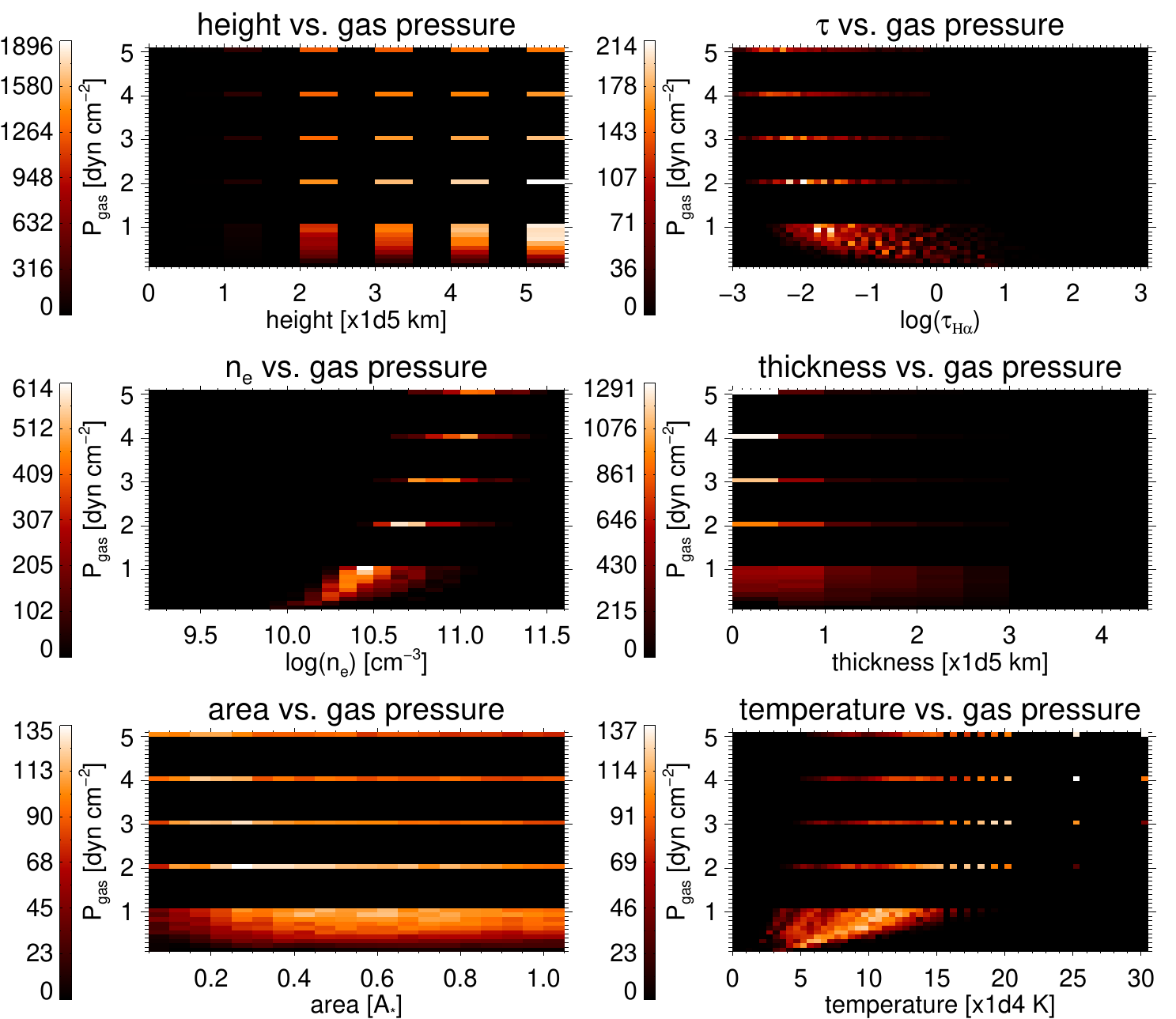}
	\includegraphics[width=7cm]{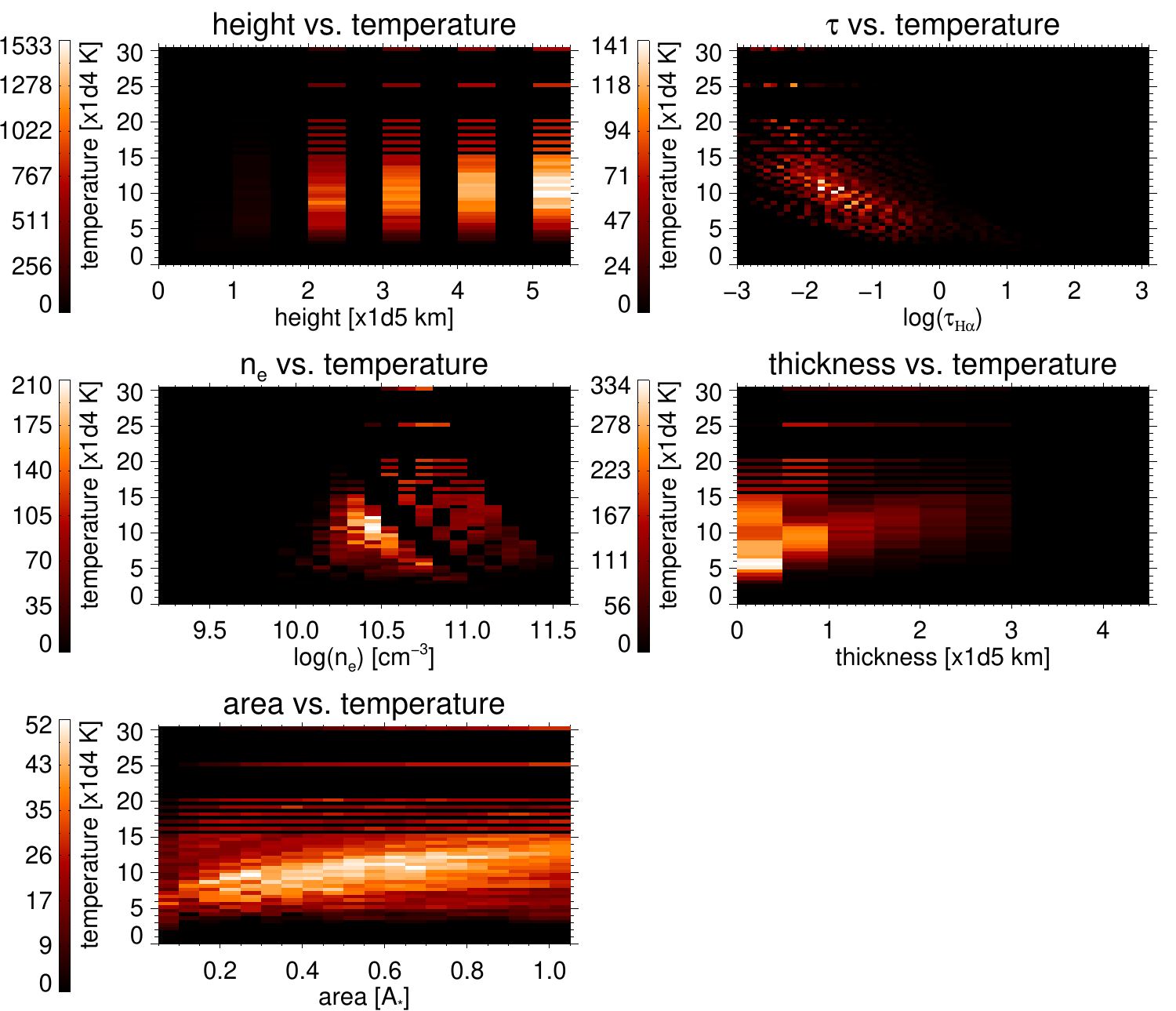}
	\includegraphics[width=7cm]{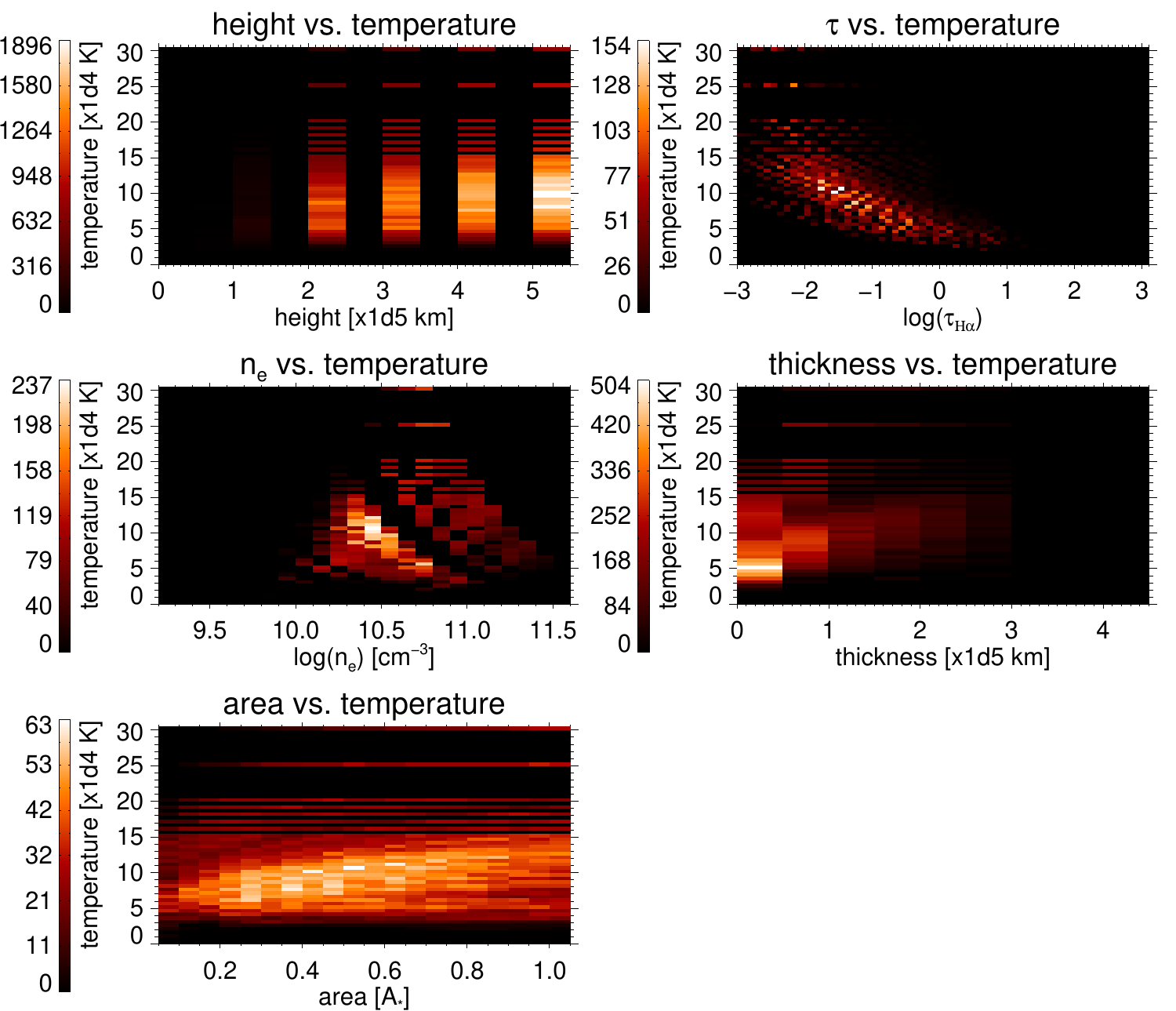}
    \caption{2D histograms of all parameter combinations of the 2-$\sigma$ cloud model results for spectrum no.~112. Columns 1 and 2: 2D histograms of all parameter combinations of the 2-$\sigma$ cloud model results for the prominence case. Columns 3 and 4: 2D histograms of all parameter combinations of the 2-$\sigma$ cloud model results for the filament case. \label{dep4}}
\end{center}
\end{figure*}



\bsp	
\label{lastpage}
\end{document}